\documentclass[12pt]{article}

\usepackage{amssymb}

\usepackage{graphicx}

\begin{document}

\title{An anisotropic Kantowski-Sachs universe with radiation, dust and a phantom fluid}

\author{G. Oliveira-Neto and D. L. Canedo\\
Departamento de F\'{\i}sica, \\
Instituto de Ci\^{e}ncias Exatas, \\ 
Universidade Federal de Juiz de Fora,\\
CEP 36036-330 - Juiz de Fora, MG, Brazil.\\
gilneto@fisica.ufjf.br, danielcanedo.tr@hotmail.com
\and G. A. Monerat\\
Departamento de Modelagem Computacional, \\
Instituto Polit\'{e}cnico, \\
Universidade do Estado do Rio de Janeiro, \\
CEP 28.625-570, Nova Friburgo - RJ - Brazil.\\
monerat@uerj.br}

\maketitle

\begin{abstract}
In the present work, we study the dynamical evolution of an homogeneous and anisotropic KS cosmological model, considering general relativity as the gravitational theory, such that there are three different perfect fluids in the matter sector. They are radiation, dust and phantom fluid. Our main motivation is determining if the present model tends to an homogeneous and isotropic FRW model, during its evolution. Also, we want to establish how the parameters and initial conditions of the model, quantitatively, influence the isotropization of the present model. In order to simplify our task, we use the Misner parametrization of the KS metric. In terms of that parametrization the KS metric has two metric functions: the scale factor $a(t)$ and $\beta(t)$, which measures the spatial anisotropy of the model. We solve, numerically, the Einstein's equations of the model and find a solution where the universe starts to expand from a, small, initial size and continues to expand until it ends in a {\it Big Rip} singularity. We explicitly show that for the expansive solution, after same time, the universe becomes isotropic. Based on that result, we can speculate that the expansive solution may represent an initial, anisotropic, stage of our Universe, that later, due to the expansion, became isotropic.
\end{abstract}






\section{Introduction}
\label{intro}

In the early moments, after its birth, the Universe may had been very different from what it is today. For instance, many physicists believe that during a brief moment, just after the birth of the Universe, the gravitational interaction, governing the dynamics of the Universe, was quantized. Therefore, the geometry of the Universe, at the beginning, must have had a foam like structure, resonating between one configuration and another and another \cite{wheeler}. If one accepts that possibility, it is natural to question the validity of the {\it Cosmological Principle}, at the early moments of the Universe. That principle states that the Universe is homogeneous and isotropic, at sufficiently large scales \cite{dinverno}. Let us suppose that the {\it Cosmological Principle} is not valid, at the beginning of the Universe.
Then, let us consider that the Universe is initially inhomogeneous and anisotropic. That initial inhomogeneous and anisotropic state, after some time, was transformed in the present homogeneous and isotropic state we know today. That transformation must had happened before the decoupling between matter and radiation, because the Cosmic Microwave Background Radiation, produced due to the decoupling, is almost uniformly distributed in all directions with very small irregularities in different directions \cite{liddle}. One moment when the isotropization
and homogenization, of the primordial Universe, could had taken place, was during the {\it Cosmological Inflation}. During that very brief period of time, just after the initial singularity, the Universe expanded in an accelerated rate and increased its size many orders of magnitude \cite{liddle1}. That idea, of an initial inhomogeneous and anisotropic Universe that later becomes homogeneous and isotropic, is not new \cite{misner0} and many physicists have already contributed to that area. Many of those contributions are concentrated in a particular situation, where the initial Universe is homogeneous and anisotropic. Even considering that particular situation, one is left with many options for the choice of a particular homogeneous and anisotropic spacetime. One very interesting candidate is the Kantowski-Sachs (KS) spacetime \cite{KS}. That spacetime has a $S^2 \times \Re$ spatial topology (or $S^2 \times S^1$, if the real line is compactified due to identifications), therefore due to the spherical symmetry one needs just two scale factors to describe it. That fact is one of the appealing properties of the KS spacetime. The curvatures of the spatial slices, of that spacetime, are constants and positives. Another important property of the KS spacetime is that it may describe the interior of a Schwarzchild black hole \cite{KS}. In the traditional parametrization the KS metric is given by,
\begin{equation}
\label{1}
ds^{2} = -dt^{2} + \bar{a}(t)^{2}dr^{2} + b(t)^{2}(d\theta^{2} + \sin^{2}\theta \, d\phi^{2}).
\end{equation}
Where $t$ is the time coordinate, $\bar{a}(t)$ e $b(t)$ are scale factors, $r$ is the radial coordinate, such that, $r \in {[0,+\infty)}$ and $\theta$ and $\phi$ are the spherical angular coordinates. They vary, respectively, in the ranges: $[0,\pi]$ and $[0,2\pi]$. We are using the natural unit system, where $c=8\pi G=1$.
Many works, considering general relativity as the gravitational theory, have already been produced where the Universe started with a KS metric and later had a period of rapid expansion. Some of these works, explicitly, discuss the isotropization due to the period of rapid expansion.  We give some examples of those works in Refs. \cite{lorenz,weber,weber1,gron,gron1,barrow,jensen,krori,byland,li,tiwari,adhav,chaubey,adhav1,parisi,keresztes}. Even if the gravitational theory is not general relativity, we may find several works where the cosmological models have a KS spacetime. We give some examples of those works in Refs. \cite{singh,barrow1,samanta,latta,dutta,vinutha,hoogen,mishra,cesare,mohandas,leon}.
Finally, we would like to mention that some authors studied quantum cosmological models such that the spacetime is a KS one. We give some examples of those works in Refs. \cite{laflamme,halliwell,compean,nelson,obregon}.

In the present work, we want to contribute to that important research area. In particular, we want to study the dynamical evolution of a KS cosmological model, considering general relativity as the gravitational theory, such that there are three different perfect fluids in the matter sector. They are: (i) radiation, which was very important at the beginning of the Universe; (ii) dust, which represents the ordinary matter of the Universe; and (iii) phantom fluid, which is the dark energy, responsible for the period of rapid expansion just after the initial
singularity. There is, also, an important contribution from dark energy to the present matter content of the Universe \cite{riess0,perlmutter}.
Our main motivation is verifying, initially, if the present model tends to an homogeneous and isotropic Friedman-Roberson-Walker (FRW) model,
during its evolution. If the isotropization of the present model, indeed, takes place, we want to determine how the parameters and initial conditions, of the model, influence it.
In order to simplify our task, we use a parametrization of the KS metric different from Eq. (\ref{1}). It is called the Misner parametrization
and was first introduced in Ref. \cite{misner}. Therefore, inspired by the Misner parametrization, we may write the KS metric in the following way,
\begin{equation} 
\label{4}
ds^{2} = -dt^{2} + a(t)^{2}e^{-\beta(t)}dr^{2} + a(t)^{2}e^{\beta(t)}(d\theta^{2} + \sin^{2}\theta \, d\phi^{2}),
\end{equation}
where $a(t)$ and $\beta(t)$ are functions of the time coordinate. One may recover the KS metric, written in the traditional parametrization 
Eq. (\ref{1}), if one imposes the following conditions on the metric functions of Eq. (\ref{4}), 
\begin{equation}
\label{1,5}
a(t) = \sqrt{\bar{a}(t) b(t)},\qquad \beta(t) = \ln\left(\frac{b(t)}{\bar{a}(t)}\right).
\end{equation}
Observing Eq. (\ref{4}), we notice that if $\beta(t)$ tends to zero or a constant value, for a given, finite or infinity, value of time, the KS metric Eq. (\ref{4}) goes to the FRW metric, for that given value of time.
Therefore, it is very simple to identify the isotropization of the KS spacetime, using the KS metric written in the Misner parametrization Eq. (\ref{4}). Based on those considerations, we may interpret the metric functions $a(t)$ and $\beta(t)$, in the following way: $a(t)$ is a scale factor and $\beta(t)$ is a function that measures the spatial anisotropy of the model. Then, in order to investigate under which conditions our anisotropic KS model goes to an isotropic FRW one, we must study the dynamical evolution of $a(t)$ and $\beta(t)$.

In Section \ref{hamiltonian}, we compute the Hamiltonian of the KS model coupled to three different perfect fluids: radiation, dust and phantom. Then, using that Hamiltonian we derive the coupled system of differential equations for the metric variables. In Section \ref{results}, we start constructing, in Subsection \ref{portraits}, phase portraits of the model, in order to give a general idea of the different dynamical behaviors of the metric functions. In Subsection \ref{solutions}, we solve, numerically, the system of coupled differential equations and determine how the scale factor $a(t)$ and the anisotropic function $\beta(t)$ behave, as functions of the time coordinate, few parameters and the initial conditions. We pay special attention for the solution representing expansion. Finally, in Section \ref{conclusions}, we give our conclusions.

\section{The Hamiltonian and Field Equations of the Model}
\label{hamiltonian}

As we have mentioned in Section \ref{intro}, we want to study an homogeneous and anisotropic cosmological model with the KS metric 
Eq. (\ref{4}). The matter content of the model is represented by a perfect fluid with four-velocity $U^\mu = \delta^{\mu}_0$, in the comoving coordinate system used. The total energy-momentum tensor is given by,
\begin{equation}
T_{\delta \nu} = (\rho+p)U_{\delta}U_{\nu} + p g_{\delta \nu}\, ,  
\label{2}
\end{equation}
where $\rho$ and $p$ are the energy density and pressure of the fluid, respectively. Here, we assume the following equation of state for the fluid, 
\begin{equation}
\label{3}
p = w \rho, 
\end{equation}
where $w$ is a constant which defines the fluid. The matter content of our model consist of three different perfect fluids,
representing some of the matter, we believe, are present in our Universe. They are: a radiation perfect fluid ($w = 1/3$), a dust
perfect fluid ($w = 0$) and a phantom perfect fluid ($w < -1$) \cite{caldwell}.

We must start, that study, by writing the Einstein's equations for the present model. In order to do that, we consider the Hamiltonian formalism. Therefore, we begin computing the action ($S$) of the model,
\begin{equation}
\label{5}
S=\int\left\{\frac{R}{2}+{\cal L}_{m}\right\}\sqrt{-g}d^4x,
\end{equation}
where $R$ is the Ricci scalar, ${\cal L}_{m}$ is the matter Lagrangian and $g$ is the determinant of the metric Eq. (\ref{4}). For the metric Eq. (\ref{4}), the Ricci scalar is given by,
\begin{equation}
\label{6}
R=\dot{\beta}^2+\ddot{\beta}+\frac{6\dot{a}^2}{a^2}+6\frac{\ddot{a}}{a}+\frac{2}{a^2 e^{\beta}}+\frac{4\dot{a}\dot{\beta}}{a}.
\end{equation}
For a perfect fluid,
\begin{equation}
\label{6.5}
{\cal L}_{m} = -\rho,
\end{equation}
where $\rho$ is the perfect fluid energy density. In order to determine $\rho$, as function of $a(t)$ and $\beta(t)$, we write
the $T_{\mu \nu}$ expression Eq. (\ref{2}), using the KS metric Eq. (\ref{4}) and the four-velocity $U^\mu$ in the comoving coordinates. We obtain the following non-zero components of the energy-momentum tensor,
\begin{eqnarray}
\label{7}
T^{t t} & = & \left( \rho + p \right)U^{t}U^{t} + pg^{t t} = \rho,\nonumber\\
T^{r r} & = & \left( \rho + p \right)U^{r}U^{r} + pg^{r r} = pa^{-2}e^{\beta},\nonumber\\
T^{\theta \theta} & = & \left( \rho + p \right)U^{\theta}U^{\theta} + pg^{\theta \theta} = pa^{-2}e^{-\beta},\nonumber\\
T^{\phi \phi} & = & \left( \rho + p \right)U^{\phi}U^{\phi} + pg^{\phi \phi} = pa^{-2}e^{-\beta}\sin^{-2}{\theta}.\nonumber\\
\end{eqnarray}
It is important to mention that, as a simplification, we consider that the perfect fluid is isotropic, in the sense that the pressure $p$ is
the same in all directions.
Now, we may compute from the energy-momentum tensor conservation equation, $T^{\mu \nu}_{\phantom{0\nu} ; \nu} = 0$, with the aid of the non-zero components of $T_{\mu \nu}$ Eqs. (\ref{7}), an expression relating the energy density $\rho(t)$ with the metric functions $a(t)$ and 
$\beta(t)$. For a perfect fluid with equation of state given by Eq. (\ref{3}),
it is given by,
\begin{equation}
\label{8}
\rho = Ca^{-3(1 + w)}e^{\frac{-\beta}{2}(1+w)},
\end{equation}
Where $C$ is a positive integration constant associated to the fluid energy density, at a given moment. As we have mentioned before, the matter content of our model consist of three different types of perfect fluids, a radiation perfect fluid ($\omega = 1/3$), a dust
perfect fluid ($\omega = 0$) and a phantom perfect fluid ($\omega < -1$). Therefore, the total energy density of the matter content of the model, which we call $\rho_c$, may be written, with the aid of Eq. (\ref{8}), by,
\begin{equation} 
\label{9}
\rho_{c} = C_{r}a^{-4}e^{-\frac{2\beta}{3}} + C_{d}a^{-3}e^{-\frac{\beta}{2}} + C_{p} a^{-3(1+w)}e^{-\frac{\beta(1+w)}{2}},
\end{equation}
where $C_{r}$, $C_{d}$ and $C_{p}$ are the integration constants for the radiation, dust and phantom fluid, respectively. 
Now, introducing the results from Eqs. (\ref{6}) and (\ref{9}) in the action (\ref{5}), we obtain, the following action,
\begin{equation}
\label{9.1}
S=\int{\cal L}_{eff}dt,
\end{equation}
where ${\cal L}_{eff}$ is the sum of the matter and gravitational Lagrangians and we discarded a numerical multiplicative constant. After
performing two integrations by part, we derive the following expression for ${\cal L}_{eff}$,
\begin{eqnarray}
\label{9.2}
\nonumber
{\cal L}_{eff}&=&
\frac{a}{2}e^{\beta/2}\left\{\frac{a^2}{2}\dot{\beta}^2-6\dot{a}^2-2a\dot{a}\dot{\beta} -2e^{-\beta}
\right\}\\ 
&-&C_{r}a^{-1}e^{\frac{-\beta}{6}}-C_{d}-C_{p}a^{-3w}e^{-\frac{\beta}{2}w}.
\label{eq11}\\ \nonumber
\end{eqnarray}
The next step in order to write the Hamiltonian of the model is to compute the canonically conjugated momenta to $a(t)$ and $\beta(t)$. They
are given by,
\begin{eqnarray}
\label{9.3}
\nonumber
p_{a}    &=& \frac{\partial {\cal L}_{eff}}{\partial \dot{a}} =\frac{a}{2}e^{\beta/2}\left\{
-12\dot{a}-2a\dot{\beta}\right\}\\ 
p_{\beta}&=& \frac{\partial {\cal L}_{eff}}{\partial \dot{\beta}} =\frac{a}{2}e^{\beta/2}\left\{
a^2\dot{\beta}-2a\dot{a}\right\}\label{momentos}.\\ \nonumber
\nonumber
\end{eqnarray}
Now, inverting Eqs. (\ref{9.3}), we may write $\dot{a}$ and $\dot{\beta}$ in terms of $p_{a}$ and $p_{\beta}$. That furnishes,
\begin{eqnarray}
\nonumber
\dot{a}     &=& -\frac{(ap_{a}+2p_{\beta})}{8a^2\, e^{\beta/2}}\\ 
\dot{\beta} &=& -\frac{(ap_{a}-6p_{\beta})}{4a^3e^{\beta/2}}. 
\label{9.4}\\ \nonumber
\end{eqnarray}
From the general expression for an Hamiltonian, we obtain, for the present model, the following Hamiltonian,
\begin{equation}
\label{9.5}
{\cal H}_{eff} = p_{a}\dot{a} + p_{\beta}\dot{\beta} - {\cal L}_{eff}.
\end{equation}
With the aid of Eqs. (\ref{9.2}) and (\ref{9.4}), the effective Hamiltonian, ${\cal H}_{eff}$ Eq. (\ref{9.5}), has the following expression,
\begin{equation}
\nonumber
{\cal H}_{eff} = \left(-\frac{p_{a}^2}{16a}+\frac{3p_{\beta}^2}{4a^3}-\frac{p_{a}p_{\beta}}{4a^2}-a\right)e^{-\frac{\beta}{2}}+\frac{C_{r}}{a}e^{-\frac{\beta}{6}}+C_{d}+C_{p}a^{-3w}e^{-\frac{\beta}{2}w}.
\label{9.6}
\end{equation}
For the present model, ${\cal H}_{eff}$ is the superhamiltonian \cite{wheeler}. If we impose the superhamiltonian constraint, or, in other words, that ${\cal H}_{eff}$ vanishes, we obtain the ($t$, $t$) component of the Einstein's equations. Therefore, if we impose that 
${\cal H}_{eff}$ Eq. (\ref{9.6}) vanishes, rewrite the momenta $p_{a}$ and $p_{\beta}$ in terms of $\dot{a}$ and $\dot{\beta}$, in the resulting equation and, finally, multiply the resulting equation by $-\frac{1}{a^3}\, e^{-\frac{\beta}{2}}$, we obtain,
\begin{equation}
\label{10}
-\frac{\dot{\beta}^{2}}{4}
+\frac{e^{-\beta}}{a^2}+
\frac{\dot{a}\dot{\beta}}{a}+
\frac{3\dot{a}^2}{a^2}
=\frac{C_{r}}{a^4} e^{-\frac{2\beta}{3}}+\frac{C_{d}}{a^3}e^{-\frac{\beta}{2}}+C_{p}a^{-3(1+w)}e^{-\frac{\beta}{2}(1+w)}
\end{equation}
Now, from the effective Hamiltonian ${\cal H}_{eff}$ Eq. (\ref{9.6}), we compute the Hamilton's equations in order to find the other, independent, Einstein's equations. The Hamilton's equations are given by,
\begin{equation}
\dot{a}=\frac{\partial {\cal H}}{\partial p_a}=\left(-\frac{p_{a}}{8a}-\frac{p_{\beta}}{4a^{2}}
\right) e^{-\frac{\beta}{2}}
\label{10.1}
\end{equation}
\begin{equation}
\dot{p}_{a}=-\frac{\partial {\cal H}}{\partial a}=-\left( 
\frac{p_{a}^{2}}{16a^2}-\frac{9p_{\beta}^{2}}{4a^4}+\frac{p_{a}p_{\beta}}{2a^3}-1\right) e^{-\frac{\beta}{2}}+
\frac{C_{r}}{a^2}e^{-\frac{\beta}{6}}+\frac{3wC_{p}}{a^{3w+1}}e^{-\frac{w\beta}{2}}
\label{10.2}
\end{equation}
\begin{equation}
\dot{\beta}=\frac{\partial {\cal H}}{\partial p_{\beta}}=\left(\frac{3p_{\beta}}{2a^3}-\frac{p_{a}}{4a^2}\right) e^{-\frac{\beta}{2}}
\label{10.3}
\end{equation}
\begin{equation}
\dot{p}_{\beta}=-\frac{\partial {\cal H}}{\partial \beta}=\frac{1}{2}\left( 
-\frac{p_{a}^{2}}{16a}+\frac{3p_{\beta}^{2}}{4a^3}-\frac{p_{a}p_{\beta}}{4a^2}-a\right) e^{-\frac{\beta}{2}}
+\frac{C_{r}}{6a}e^{-\frac{\beta}{6}}+\frac{wC_{p}}{2a^{3w}}e^{-\frac{w\beta}{2}}
\label{10.4}
\end{equation}
Next, we must work with those equations in order to obtain the independent Einstein's equations. We start computing the second derivative of Eq. (\ref{10.1}). Then, we introduce, in the resulting equation, the values of $\dot{p}_{a}$ Eq. (\ref{10.2}), $\dot{p}_{\beta}$ Eq. (\ref{10.4}), $p_a$ and $p_\beta$ Eqs. (\ref{9.3}). Finally, we multiply it by $-2a$ and obtain,
\begin{equation}
-2a\ddot{a}-\dot{a}^2-\frac{a^2\dot{\beta}^2}{4}=\frac{C_{r}}{3a^2}e^{-\frac{2\beta}{3}}+\frac{wC_{p}}{a^{3w+1}}e^{-\frac{\beta}{2}(w+1)}
\label{10.5}
\end{equation}
Now, we compute the second derivative of Eq. (\ref{10.3}). Then, we introduce, in the resulting equation, the values of $\dot{p}_{a}$ Eq. (\ref{10.2}), $\dot{p}_{\beta}$ Eq. (\ref{10.4}), $p_a$ and $p_\beta$ Eqs. (\ref{9.3}). Finally, we multiply it by $-a^2$ and obtain,
\begin{equation}
-a^2\ddot{\beta}=3a\dot{a}\dot{\beta}+e^{-\beta}+a^2\frac{\dot{\beta}^2}{2}.
\label{10.6}
\end{equation}
If we sum Eqs. (\ref{10.5}) and (\ref{10.6}) and rewrite the RHS of the resulting equation, such that it may be written in terms of the pressure $p_c$ associated to $\rho_c$ Eq. (\ref{9}), we find the following equation, 
\begin{equation}
\label{11}
\frac{-3a^{2}\dot{\beta}^{2}}{4} - 3a\dot{a}\dot{\beta} - a^{2}\ddot{\beta} - \dot{a}^{2} - 2a\ddot{a} - e^{-\beta} = p_c a^{2},
\end{equation}
where the pressure $p_c$, with the aid of Eqs. (\ref{3}) and (\ref{9}) is given, by,
\begin{equation}
\label{12}
p_c = \frac{1}{3} C_{r}a^{-4}e^{-\frac{2\beta}{3}} + w C_{p} a^{-3(1+w)}e^{-\frac{\beta(1+w)}{2}}.
\end{equation}
Eq. (\ref{11}) is the ($r$, $r$) component of the Einstein's equations. Finally, the final, independent, Einstein's equation is obtained
by rewrite the RHS of Eq. (\ref{10.5}), such that it may be written in terms of the pressure $p_c$ Eq. (\ref{12}). Then, we find,
\begin{equation}
\label{13}
\frac{-a^{2}\dot{\beta}^{2}}{4} - \dot{a}^{2} - 2a\ddot{a} = p_c a^{2}.
\end{equation}
Eq. (\ref{13}) is the ($\theta$, $\theta$) or ($\phi$, $\phi$) component of the Einstein's equations, because they are identical in the present model.

Eqs. (\ref{10}), (\ref{12}) and (\ref{13}) form a system of second order, ordinary differential equations. In order to solve it, we must furnish initial conditions for $a(t)$ ($a_0$), $\beta(t)$ ($\beta_0$), $\dot{a}(t)$ ($\dot{a}_0$) and $\dot{\beta}(t)$ ($\dot{\beta}_0$). Observing the system, Eqs. (\ref{10}), (\ref{12}) and (\ref{13}), we notice that there are three equations and two variables: $a(t)$ and $\beta(t)$ . Therefore, we need only two equations in order to compute the dynamical evolution of those variables. We decided to use Eq. (\ref{10}) along with one of the remaining equations or a combinations of those equations. After several tests, showing that, numerically, any choice would lead to the same qualitative behavior for the dynamical evolutions of $a(t)$ and $\beta(t)$, with very small quantitative differences, we have decided to use Eq. (\ref{10}) along with the linear combination: (\ref{13}) - (\ref{11}),
\begin{equation} 
\label{14}
\frac{a^{2}\dot{\beta}^{2}}{2} + 3a\dot{a}\dot{\beta} + a^{2}\ddot{\beta} + e^{-\beta} = 0
\end{equation}
We use the first order, ordinary differential equation (\ref{10}), in order to compute the physically acceptable initial conditions. Given the values of $a_0$, $\beta_0$, $\dot{a}_0$, and all the parameters $C_{r}$, $C_{d}$, $C_{p}$, $w$, using equation (\ref{10}), we compute the 
value of $\dot{\beta}_0$. 

\section{Results}
\label{results}

\subsection{Phase Portraits}
\label{portraits}

\begin{figure}
\begin{center}
\includegraphics[height=7.5cm,width=9.5cm]{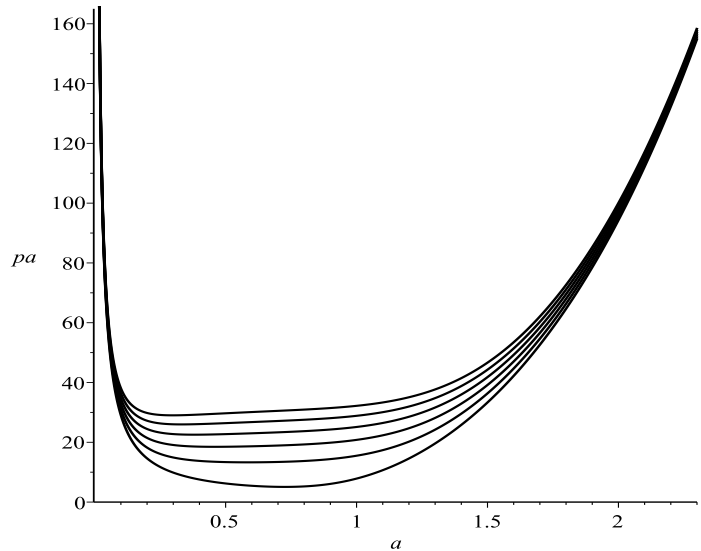}
\end{center}
\caption{Phase portrait in the plane ($a, p_a$) for different values of $C_r = 0, 10, 20, 30, 40, 50$ with $w = -2$, $C_d = C_p = 1$, $\beta = 1$, $p_{\beta} = 1.454002981$.}
\label{figure_a_portrait_expansive}
\end{figure}

\begin{figure}
\begin{center}
\includegraphics[height=7.5cm,width=9.5cm]{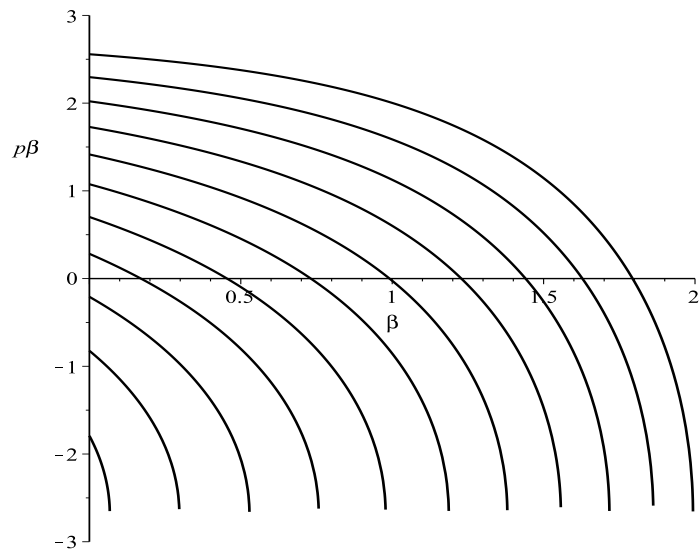}
\end{center}
\caption{Phase portrait in the plane ($\beta, p_\beta$) for different values of $C_r = 2, 4, 6, 8, 10, 12, 14, 16, 18, 20$ with $w = -2$, $C_d = C_p = 1$, $a = 1$, $p_a = -16.09777613$.}
\label{figure_b_portrait_expansive}
\end{figure}

\begin{figure}
\begin{center}
\includegraphics[height=7.5cm,width=9.5cm]{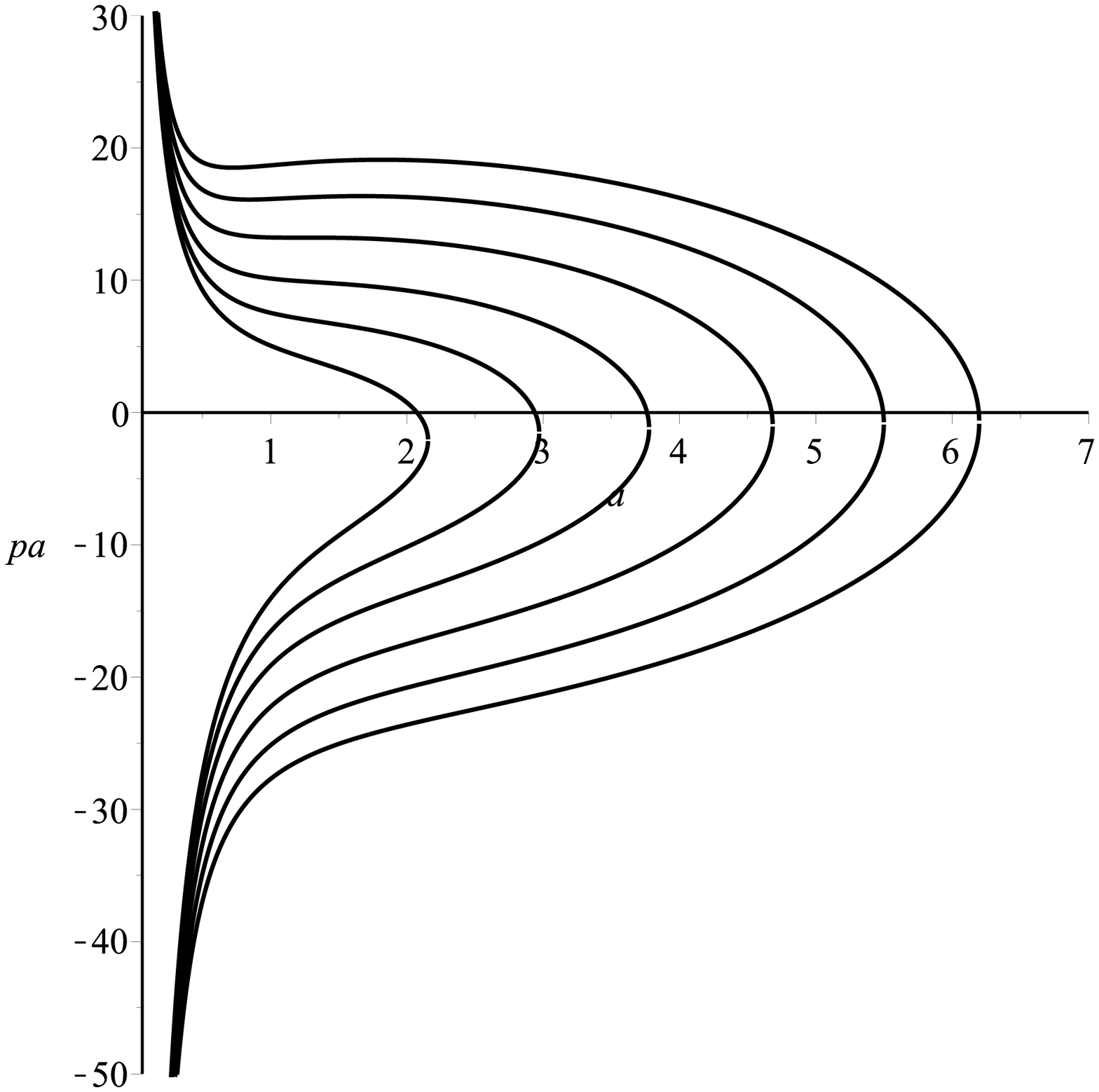}
\end{center}
\caption{Phase portrait in the plane ($a, p_a$) for different values of $C_r = 0, 2.4, 5.5, 10, 15, 20$ with $w = -2$, $C_d = 1$ and $C_p = 0.0000001$, $\beta = 1$, $p_\beta = 2.249586545$.}
\label{figure_a_portrait_bounded}
\end{figure}

Before we proceed to the detailed solution to the system of second order, ordinary differential equations formed by Eqs. (\ref{10}) and (\ref{14}), let us try to learn the general behavior of those solutions. In order to do that, we draw the phase portraits for the model, with the aid of ${\cal H}_{eff}$ Eq. (\ref{9.6}). We start imposing the superhamiltonian constraint. Then, we study, separately, the solutions to the model in the two planes ($a, p_a$) and ($\beta, p_\beta$). After considering many different values of the parameters: $w$, $C_r$, $C_d$, $C_p$; and many different, physically acceptable, values of the initial conditions: $a_0$, $\dot{a}_0$, $\beta$, $\dot{\beta}_0$; we notice that the solutions are of two types. In the first type, the scale factor $a$ starts expanding, very rapidly, from a singularity, at $a=0$. Then, it reduces the rate of expansion, for a small $a$ interval, and, finally, for larger values of $a$, it resumes the expansion, in an accelerated rate. On the other hand, the anisotropy parameter $\beta$ start expanding from $\beta=0$ and rapidly tends to a constant, finite, positive value. Therefore, in that type of solution, we have the isotropization of the model. Since, we are interested in the behavior of the solution for large values of $a$, where the isotropization takes place, we shall not study, in the rest of the paper, that solution near the initial singularity. As an example of that type of solution, we show Figures \ref{figure_a_portrait_expansive} and \ref{figure_b_portrait_expansive}. In order to draw those figures, we considered few models with different values of $C_r$, and the other parameters with the following fixed values: $w=-2$, $C_d=1$, $C_p=1$. In the other type of solution, both the scale factor $a$ and the anisotropy parameter $\beta$ start expanding, very rapidly, from a initial singularity, at the zero value. Then, they reduce the rate of expansion until they reach maximum values and stop the expansion. Then, they start a contraction until they reach a final singularity at the zero value. As an example of that type of solution, we show Figure \ref{figure_a_portrait_bounded}. In order to draw that figure, we considered few models with different values of $C_r$, and the other parameters with the following fixed values: $w=-2$, $C_d=1$, $C_p=0.0000001$. Since, in the second type of solution, there is no isotropization of the model, because $\beta$ does not go to a finite constant value or a non-singular zero value, we shall not consider that type of solution, in the rest of the paper.

\subsection{Detailed Study of the Solutions}
\label{solutions}

Now, we want to solve the system of second order, ordinary differential equations formed by Eqs. (\ref{10}) and (\ref{14}). Unfortunately,
it is not possible to find algebraic solutions to $a(t)$ and $\beta(t)$. Therefore, we solve it numerically. After we do that for many different values of the initial conditions, $a_0$, $\beta_0$, $\dot{a}_0$, $\dot{\beta}_0$, and the parameters, $w$, $C_{r}$, $C_{d}$, 
$C_{p}$, all compatible with the first type of solution described in the previous Subsection \ref{portraits}, we obtain that the scale factor $a(t)$ is expansive and $\beta(t)$ goes, asymptotically, to a constant. Then, qualitatively, $a(t)$ starts to expand from a small finite value and after a finite time interval it reaches an infinite value giving rise to a \textit{Big Rip} singularity. The presence of that singularity could not be identified in the phase portrait, because only when one solves the dynamical equations, to find $a(t)$ as a function of time, one obtains that, after a finite time interval, $a(t)$ tends to an infinity value. In the present model, we identified the \textit{Big Rip}, numerically. On the other hand, $\beta(t)$ starts to expand from a small finite value and then it goes, asymptotically, to a constant, when the universe goes to the \textit{Big Rip} singularity. The fact that, $\beta(t)$ goes, asymptotically, to a constant is very important because, as we have mentioned before, it guarantees that the solution is asymptotically isotropic. More precisely, the behavior of $\beta(t)$ means that the KS metric (\ref{4}), goes, asymptotically, to a FRW metric, where $a(t)$ plays the role of the scale factor.

Next, we investigate how the variation of the parameters, $C_{r}$, $C_{d}$, $C_{p}$, $w$, and the initial conditions, $a_0$, $\beta_0$, $\dot{a}_0$, $\dot{\beta}_0$, modify, quantitatively, the dynamical evolution of $a(t)$ and $\beta(t)$. In order to do that, we vary the value of one
of the parameters or initial conditions and we fix the values of all other quantities. In all the examples we give in the next Subsections, the values of the parameters and initial conditions are chosen for a better visualization of the results. In particular, when we are not varying
the initial conditions, they have the following values,
\begin{equation}
\label{15}
a_0 = 1,\qquad \beta_0 = 1,\qquad \dot{a}_0 = 1.
\end{equation}
In order to compute the physically acceptable initial conditions, for each case, $\dot{\beta}_0$ is free to vary. Its value is determined by the first order, ordinary differential equation (\ref{10}).

\subsection{Varying $w$}

\begin{figure}
\begin{center}
\includegraphics[height=7.5cm,width=9.5cm]{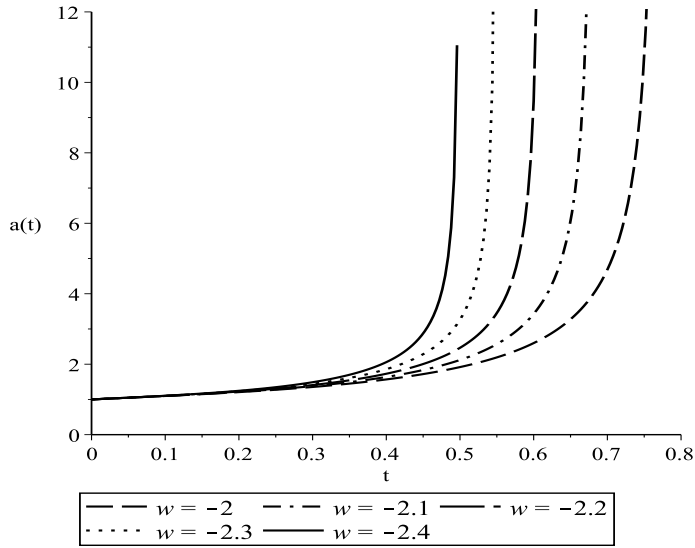}
\end{center}
\caption{$a(t)$ as a function of $t$ for different values of $w$ with $C_r = C_d = C_p = 1$}
\label{figure_aw}
\end{figure}

\begin{figure}
\begin{center}
\includegraphics[height=7.5cm,width=9.5cm]{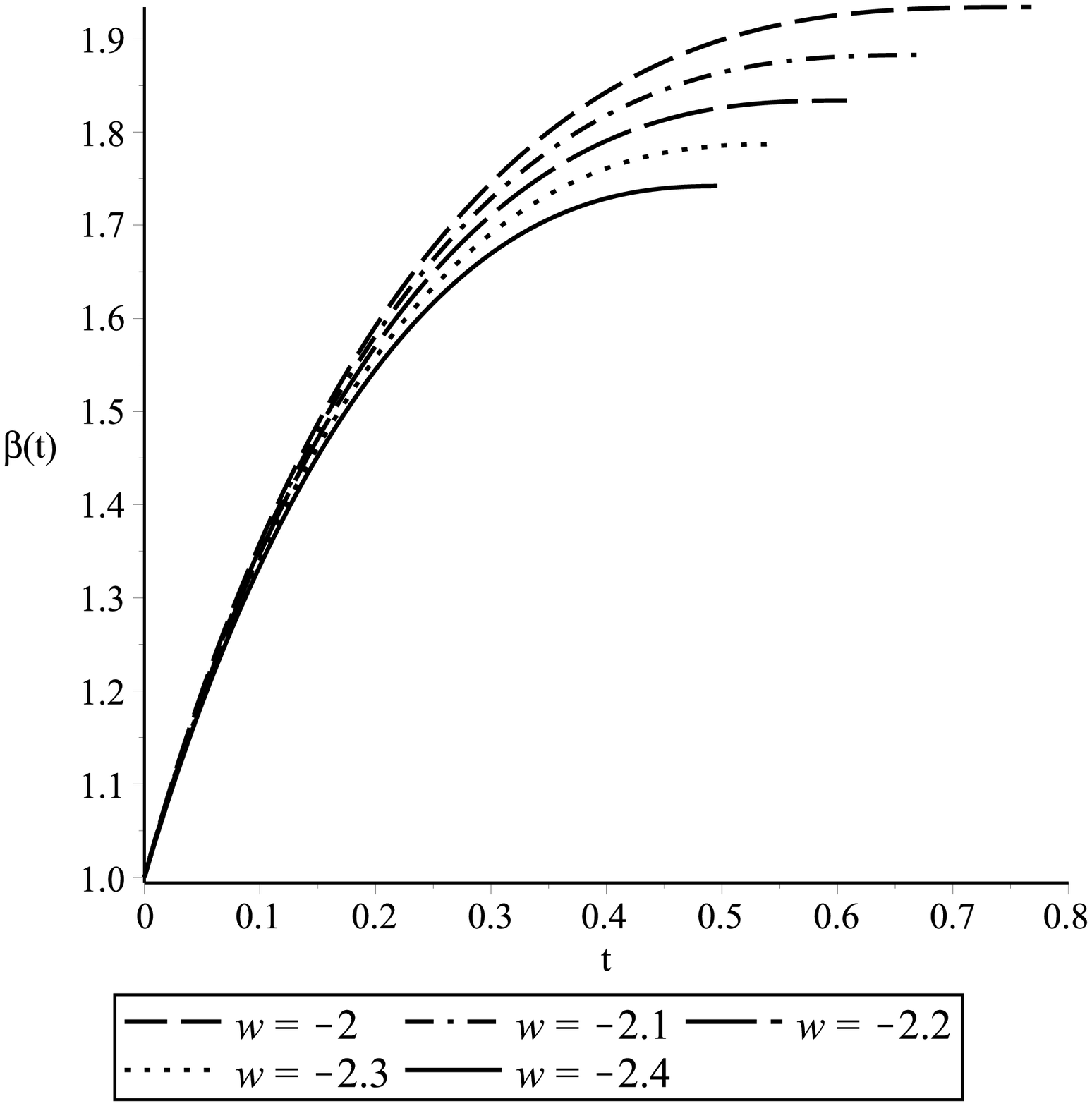}
\end{center}
\caption{$\beta(t)$ as a function of $t$ for different values of $w$ with $C_r = C_d = C_p = 1$}
\label{figure_betaw}
\end{figure}

Let us start studying how the phantom fluid parameter $w$ modifies, quantitatively, the dynamical evolution of $a(t)$ and $\beta(t)$. After computing the solution to the system Eqs. (\ref{10}), (\ref{14}), for many different values of $w$ with fixed values of $C_r$, $C_d$
and $C_p$, we find that: the scale factor $a(t)$ expands more rapidly for smaller values of $w$ and the anisotropy parameter $\beta(t)$ tends, asymptotically, to greater constant values when one increases the value of $w$. 
The fact that $a(t)$ expands more rapidly for smaller values of $w$ is expected, because in that situation the phantom fluid is becoming more, gravitationally, repulsive.
We present, respectively, in Figures \ref{figure_aw} and \ref{figure_betaw} examples of these behaviors. In order to verify that the anisotropy parameter $\beta(t)$ tends, asymptotically, to a constant value, we construct Table \ref{T1}. There, one can see the values of $t_s$, which is the time just before the universe reaches the {\it Big Rip} singularity. One can, also, see $a(t)$, $\beta(t)$, $\dot{a}(t)$, $\dot{\beta}(t)$ at $t = t_s$, for the values of $w$ shown in Figures \ref{figure_aw} and \ref{figure_betaw}. From that Table, it is clear that $\beta(t)$ tends, asymptotically, to a constant value, because its time derivative tends, asymptotically, to zero.

\begin{table}[!htb]
\centering
\caption{\footnotesize Values of $\beta(t)$, $\dot{\beta}(t)$, $a(t)$ and $\dot{a}(t)$ at $t = t_s$, for
$C_r = C_d = C_p = 1$ and different values of $w$.}
\label{T1}
{\scriptsize\begin{tabular}{c c c c c c}
\hline
$w$ &  $t_{s}$ & $\beta(t_{s})$ & $\dot{\beta}(t_{s}) \times 10^{-14}$ & $a(t_{s})$ & $\dot{a}(t_{s}) \times 10^{12}$ \\
\hline

 -2.4 & 0.49888534 & 1.74196405189664 & $84.4075472560425$ & 14840.1826703332 & $9.07162233663117$ \\

 -2.3 & 0.54956383 & 1.78693403545568 & $2.21571894047946$ & 51237.3460207391 & $80.7119785545053$ \\

 -2.2 & 0.60968911 & 1.83376486126677 & $13.5592545030986$ & 29107.5441964596 & $3.15917185779594$ \\

 -2.1  & 0.68197230 & 1.88272072974607 & $3.45298354819359$ & 73440.9146056974 & $7.60379818148587$ \\

 -2.0  & 0.77024456 & 1.93410385073669 & $3.41056094076826$ & 211224.939941673 & $19.1997256990363$ \\

\hline
\end{tabular}}
\end{table}

\subsection{Varying $C_r$ or $C_d$}

\begin{figure}
\begin{center}
\includegraphics[height=7.5cm,width=9.5cm]{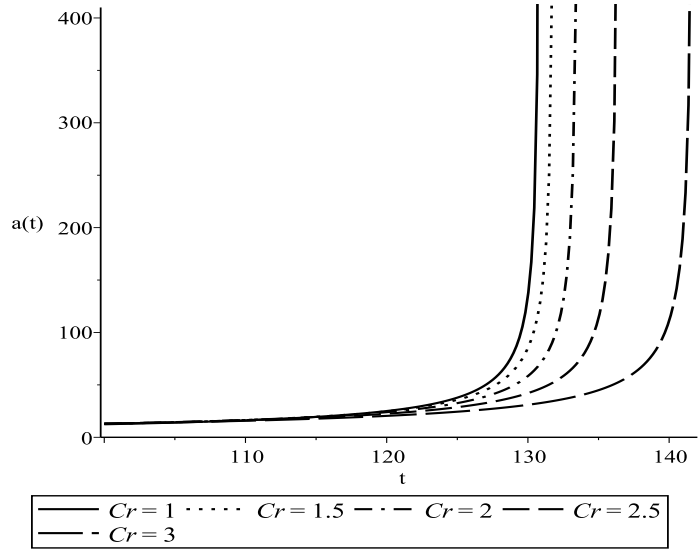}
\end{center}
\caption{$a(t)$ as a function of $t$ for different values of $C_r$ with $w = -2$, $C_d = 1$, $C_p = 0.0000001$}
\label{figure_aCr}
\end{figure}

\begin{figure}
\begin{center}
\includegraphics[height=7.5cm,width=9.5cm]{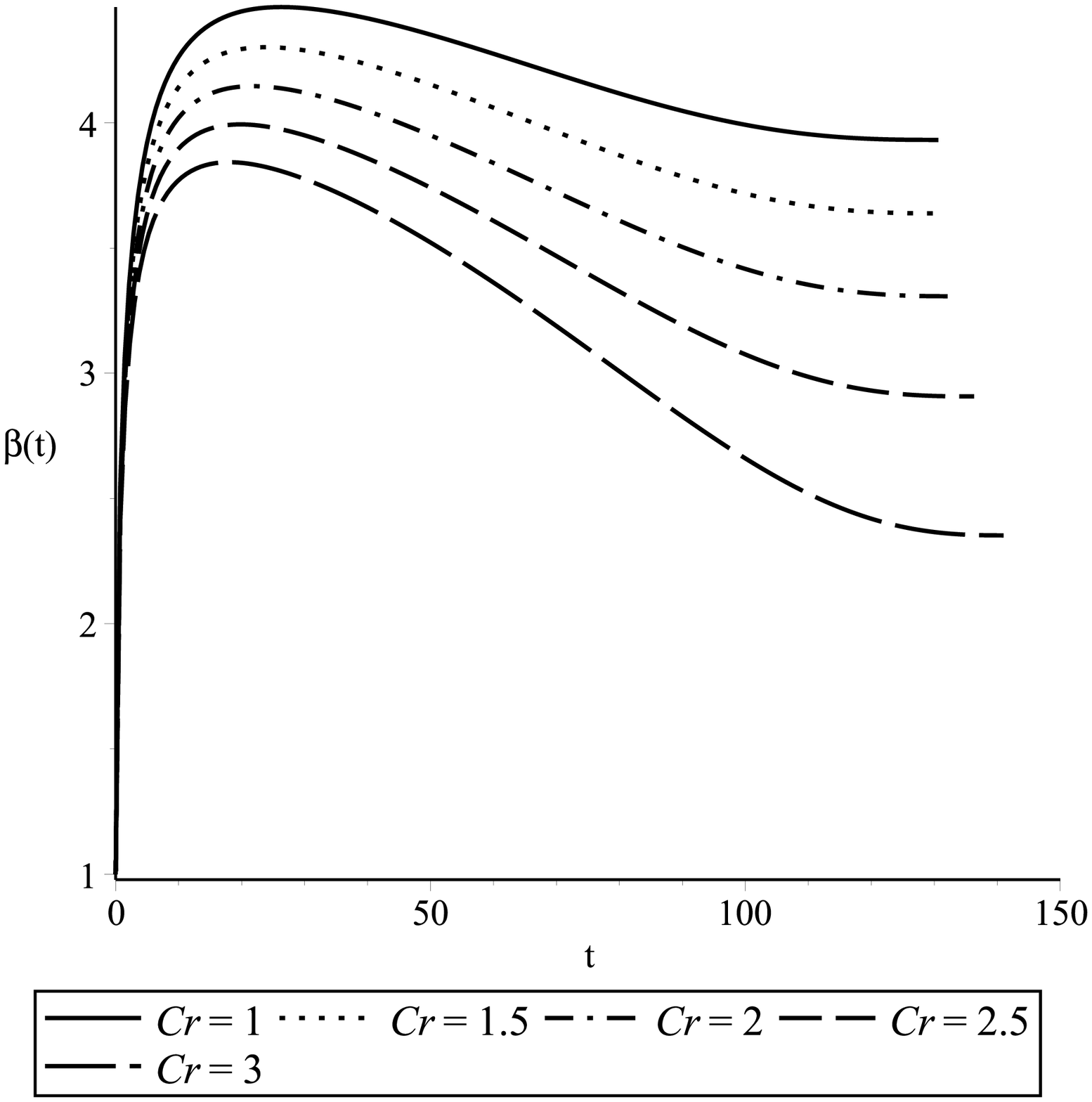}
\end{center}
\caption{$\beta(t)$ as a function of $t$ for different values of $C_r$ with $w = -2$, $C_d = 1$, $C_p = 0.0000001$}
\label{figure_betaCr}
\end{figure}

\begin{figure}
\begin{center}
\includegraphics[height=7.5cm,width=9.5cm]{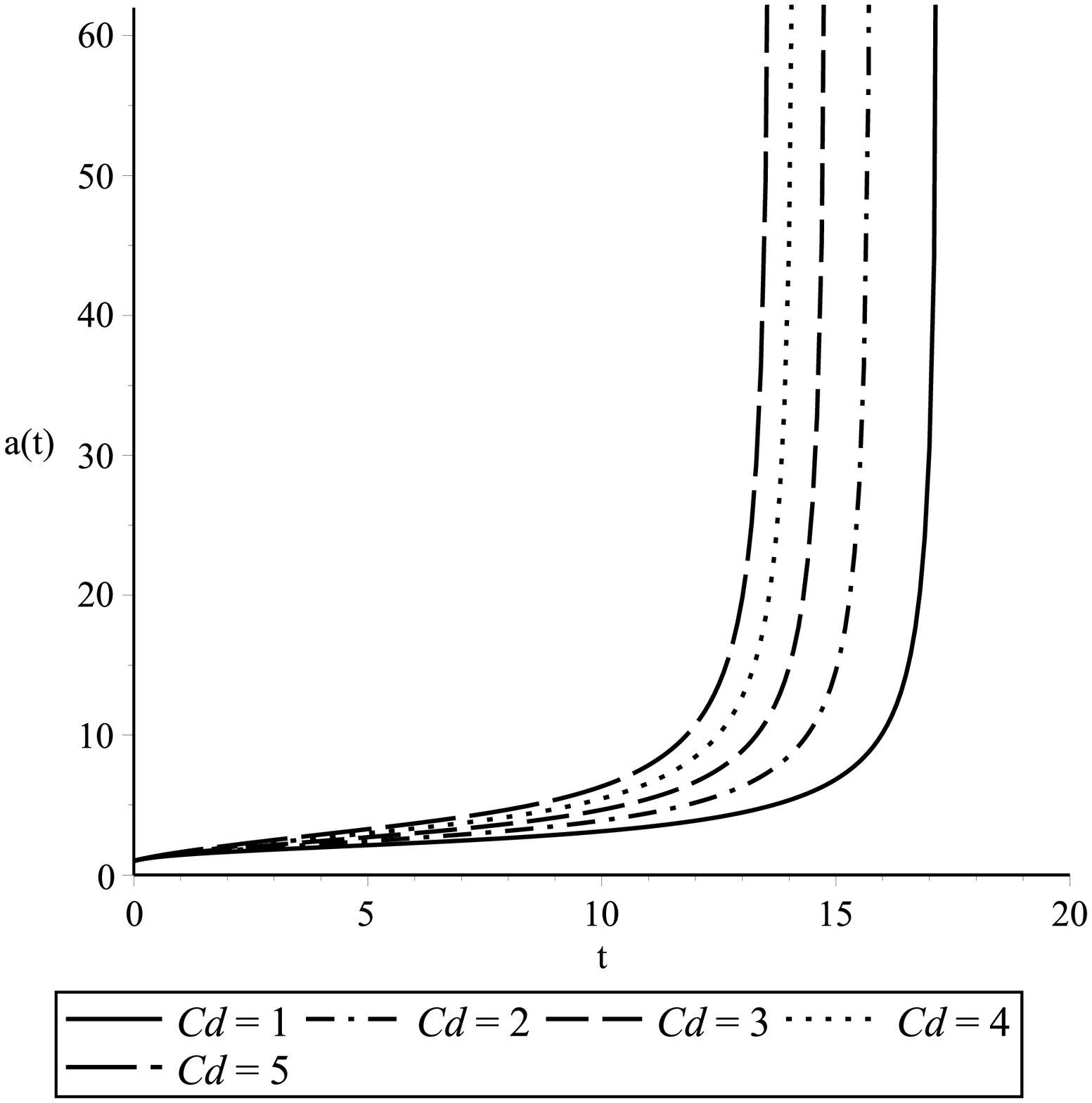}
\end{center}
\caption{$a(t)$ as a function of $t$ for different values of $C_d$ with $w = -2$, $C_r = 1$, $C_p = 0.0001$}
\label{figure_aCd}
\end{figure}

\begin{figure}
\begin{center}
\includegraphics[height=7.5cm,width=9.5cm]{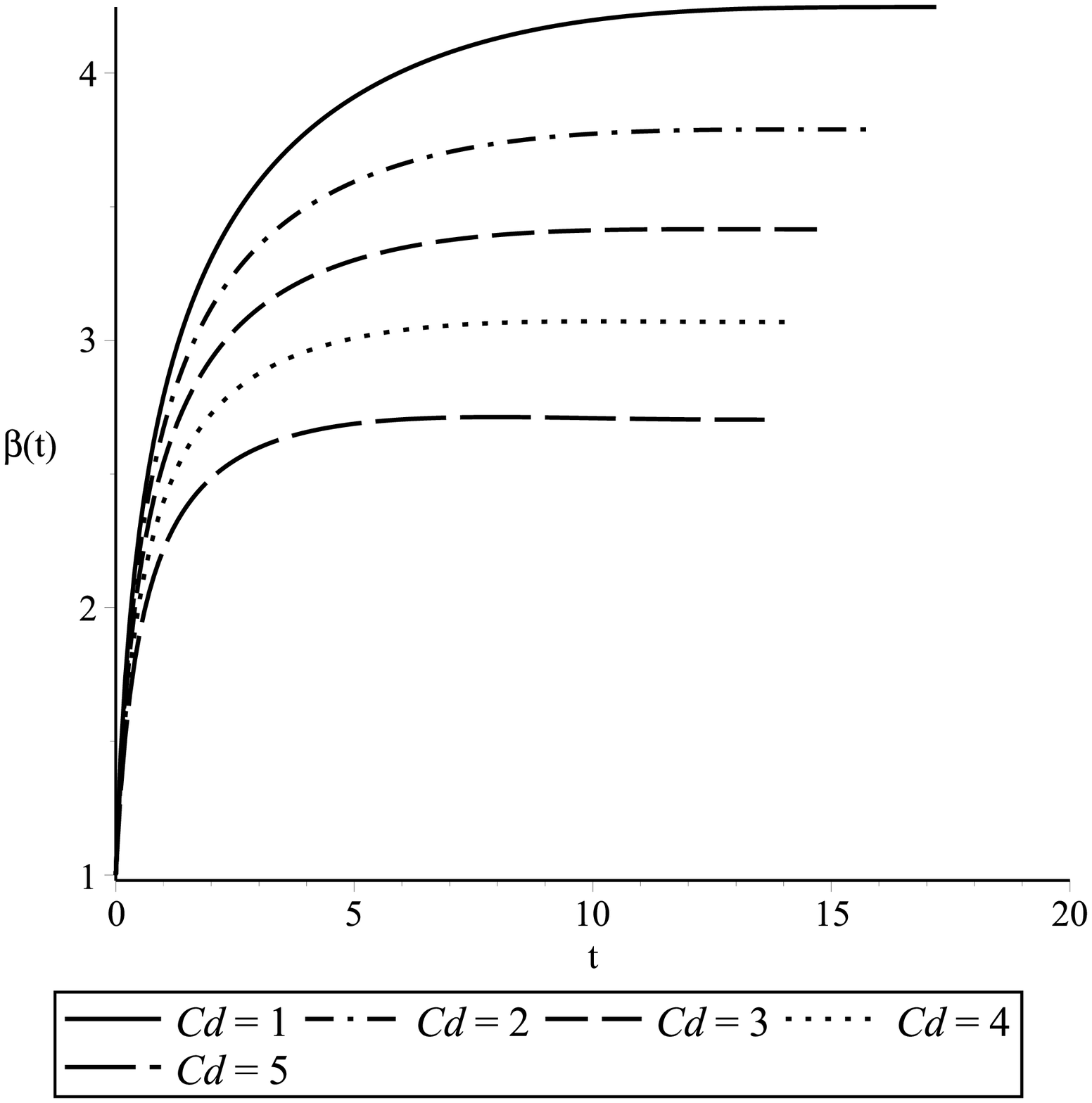}
\end{center}
\caption{$\beta(t)$ as a function of $t$ for different values of $C_d$ with $w = -2$, $C_r = 1$, $C_p = 0.0001$}
\label{figure_betaCd}
\end{figure}

Now, we may study how the variations of the energy density parameters $C_r$ or $C_d$ modify, quantitatively, the dynamical evolution of $a(t)$ and $\beta(t)$. We may study both parameters together, because their variations lead to the same general results. After computing the solution to the system Eqs. (\ref{10}), (\ref{14}), for many different values of $C_r$ ($C_d$), with fixed values of $w$, $C_d$ ($C_r$) and $C_p$, we find that: the scale factor $a(t)$ may expand more rapidly or slowly for increasing values of $C_r$ ($C_p$). It expands more slowly in two different situations: (i) when the values of $C_r$, $C_d$ and $C_p$ have the same order of magnitude; and (ii) when $C_p$ is very small in its own and also much smaller than $C_r$ and $C_d$. For all other situations $a(t)$ expands more rapidly, for increasing values of $C_r$ ($C_d$).
The anisotropy parameter $\beta(t)$ tends, asymptotically, to smaller constant values when one increases the value of $C_r$ ($C_d$). In Figures \ref{figure_aCr} and \ref{figure_betaCr}, we show examples that increasing the value of $C_r$, $a(t)$ expands more slowly and $\beta(t)$ tends, asymptotically, to smaller constant values. In Figures \ref{figure_aCd} and \ref{figure_betaCd}, we show examples that increasing the value of $C_d$, $a(t)$ expands more rapidly and $\beta(t)$ tends, asymptotically, to smaller constant values.
In Tables \ref{T2} and \ref{T3}, we show the values of $t_s$ and $a(t)$, $\beta(t)$, $\dot{a}(t)$, $\dot{\beta}(t)$ at $t = t_s$, for the values of $C_r$ and $C_p$ displayed in Figures \ref{figure_aCr}, \ref{figure_betaCr} and \ref{figure_aCd}, \ref{figure_betaCd}, respectively.
From those Tables, it is clear that $\beta(t)$ tends, asymptotically, to a constant value, because its time derivative tends, asymptotically, to zero. Although the variations of $C_r$ and $C_d$ lead to the same general results, concerning the dynamical evolutions of $a(t)$ and $\beta(t)$, it is possible to identify differences between those two parameters. For models where $a(t)$ expands more slowly, for increasing values of $C_r$ or $C_d$, $a(t)$ expands more slowly when one increases $C_r$ rather than $C_d$. We give an example of that behavior in Table \ref{T31}. There, we compare the values of $t_s$ for two models. In the first one, we vary $C_r$, leaving the other quantities fixed. In the second one, we vary $C_d$, leaving the other quantities fixed. We use the same values of all quantities in both models, in order to facilitate the comparison. For models where $a(t)$ expands more rapidly, for increasing values of $C_r$ or $C_d$, $a(t)$ expands more rapidly when one increases $C_d$ rather than $C_r$. We give an example of that behavior in Table \ref{T32}. There, we compare the values of $t_s$ for two models. In the first one, we vary $C_r$, leaving the other quantities fixed. In the second one, we vary $C_d$, leaving the other quantities fixed. We use the same values of all quantities in both models, in order to facilitate the comparison. Taking in account the results of that
comparison, between $C_r$ and $C_d$, we may say that the radiation fluid is more, gravitationally, attractive than the dust one.

\begin{table}[!htb]
\centering
\caption{\footnotesize Values of $\beta(t)$, $\dot{\beta}(t)$, $a(t)$ and $\dot{a}(t)$ at $t = t_s$, for
$C_d = 1$, $C_p = 0.0000001$, $w = -2$ and different values of $C_r$.}
\label{T2}
{\scriptsize\begin{tabular}{c c c c c c}
\hline
$C_r$ &  $t_{s}$ & $\beta(t_{s})$ & $\dot{\beta}(t_{s}) \times 10^{-11}$ & $a(t_{s})$ & $\dot{a}(t_{s}) \times 10^{10}$ \\
\hline

 1.0 & 130.87212 & 3.93224223170835 & $2.81487462114303$ & 359174.023141133 & $3.77258931834407$ \\

 1.5 & 131.87085 & 3.63852080064442 & $1.20220119374832$ & 665719.588236944 & $16.3951414888867$ \\

 2.0 & 133.57988 & 3.30768938204340 & $2.92980483031589$ & 389322.488449481 & $3.94767179235521$ \\

 2.5  & 136.45690 & 2.90764378786510 & $1.73784373547571$ & 603638.497018952 & $10.6923980123743$ \\

 3.0  & 141.72936 & 2.35254128365355 & $0.844917721724429$ & 1034268.30605848 & $35.7640315183190$ \\

\hline
\end{tabular}}
\end{table}

\begin{table}[!htb]
\centering
\caption{\footnotesize Values of $\beta(t)$, $\dot{\beta}(t)$, $a(t)$ and $\dot{a}(t)$ at $t = t_s$, for
$C_r = 1$, $C_p = 0.0001$, $w = -2$ and different values of $C_d$.}
\label{T3}
{\scriptsize\begin{tabular}{c c c c c c}
\hline
$C_d$ &  $t_{s}$ & $\beta(t_{s})$ & $\dot{\beta}(t_{s}) \times 10^{-13}$ & $a(t_{s})$ & $\dot{a}(t_{s}) \times 10^{11}$ \\
\hline

 1.0 & 17.236542 & 4.24714293260463 & $2.55148791589835$ & 793419.119050158 & $93.6103343315701$ \\

 2.0 & 15.804132 & 3.78933842465641 & $22.9274411809584$ & 195189.420349461 & $2.50612280882553$ \\

 3.0 & 14.860818 & 3.41546916176941 & $41.3644252181807$ & 144670.114660462 & $1.07948420236074$ \\

 4.0 & 14.178391 & 3.06859090847191 & $7.54569914556239$ & 472969.348275445 & $19.1288958949350$ \\

 5.0 & 13.667964 & 2.70313952681470 & $17.1823742436583$ & 281602.901740459 & $4.77553453014156$ \\

\hline
\end{tabular}}
\end{table}

\begin{table}[!htb]
\centering
\caption{\footnotesize Comparison between $C_r$ and $C_d$ for models where $a(t)$ expands more slowly for increasing values of $C_r$ or 
$C_d$. We show values of $t_s$, for two models. In the first model, one has $C_d = 1$, $C_p = 1$, $w = -2$ and different values of $C_r$. In the second model, one has $C_r = 1$, $C_p = 1$, $w = -2$ and different values of $C_d$.}
\label{T31}
{\scriptsize\begin{tabular}{c c c }
\hline
$C_{r}/C_d$ & $t_{s}(C_r)$ & $t_{s}(C_d)$\\
\hline
			
1.0  &  0.77024456 & 0.77024456\\
			
1.4  &  0.77196314 & 0.77153379\\
			
1.8  &  0.77388985 & 0.77311172\\
			
2.2  &  0.77608473 & 0.77508516\\

2.6  &  0.77864213 & 0.77764558\\
			
\hline
\end{tabular}}
\end{table}

\begin{table}[!htb]
\centering
\caption{\footnotesize Comparison between $C_r$ and $C_d$ for models where $a(t)$ expands more rapidly for increasing values of $C_r$ or 
$C_d$. We show values of $t_s$, for two models. In the first model, one has $C_d = 1$, $C_p = 0.0001$, $w = -2$ and different values of $C_r$. In the second model, one has $C_r = 1$, $C_p = 0.0001$, $w = -2$ and different values of $C_d$.}
\label{T32}
{\scriptsize\begin{tabular}{c c c }
\hline
$C_{r}/C_d$ & $t_{s}(C_r)$ & $t_{s}(C_d)$\\
\hline
			
1.0  &  17.236542 & 17.236542\\
			
2.0  &  16.929397 & 15.804132\\
			
3.0  &  16.672140 & 14.860818\\
			
4.0  &  16.462634 & 14.178391\\

5.0  &  16.306117 & 13.667964\\
			
\hline
\end{tabular}}
\end{table}

\subsection{Varying $C_p$}

\begin{figure}
\begin{center}
\includegraphics[height=7.5cm,width=9.5cm]{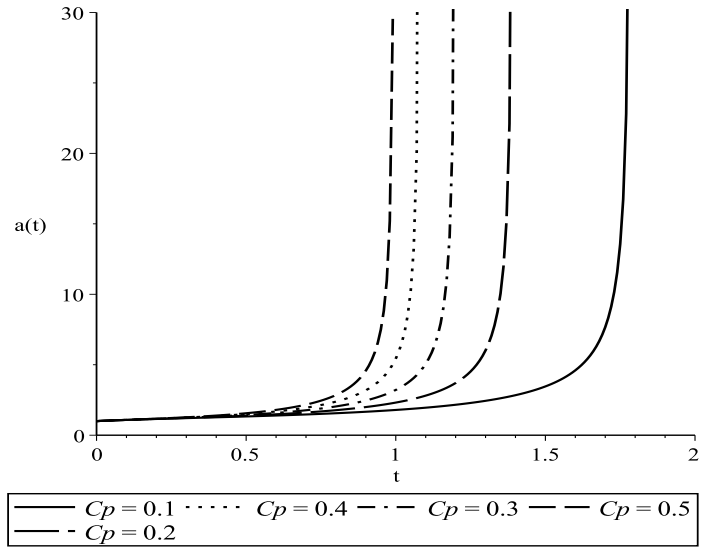}
\end{center}
\caption{$a(t)$ as a function of $t$ for different values of $C_p$ with $w = -2$, $C_r = C_d = 1$}
\label{figure_aCp}
\end{figure}

\begin{figure}
\begin{center}
\includegraphics[height=7.5cm,width=9.5cm]{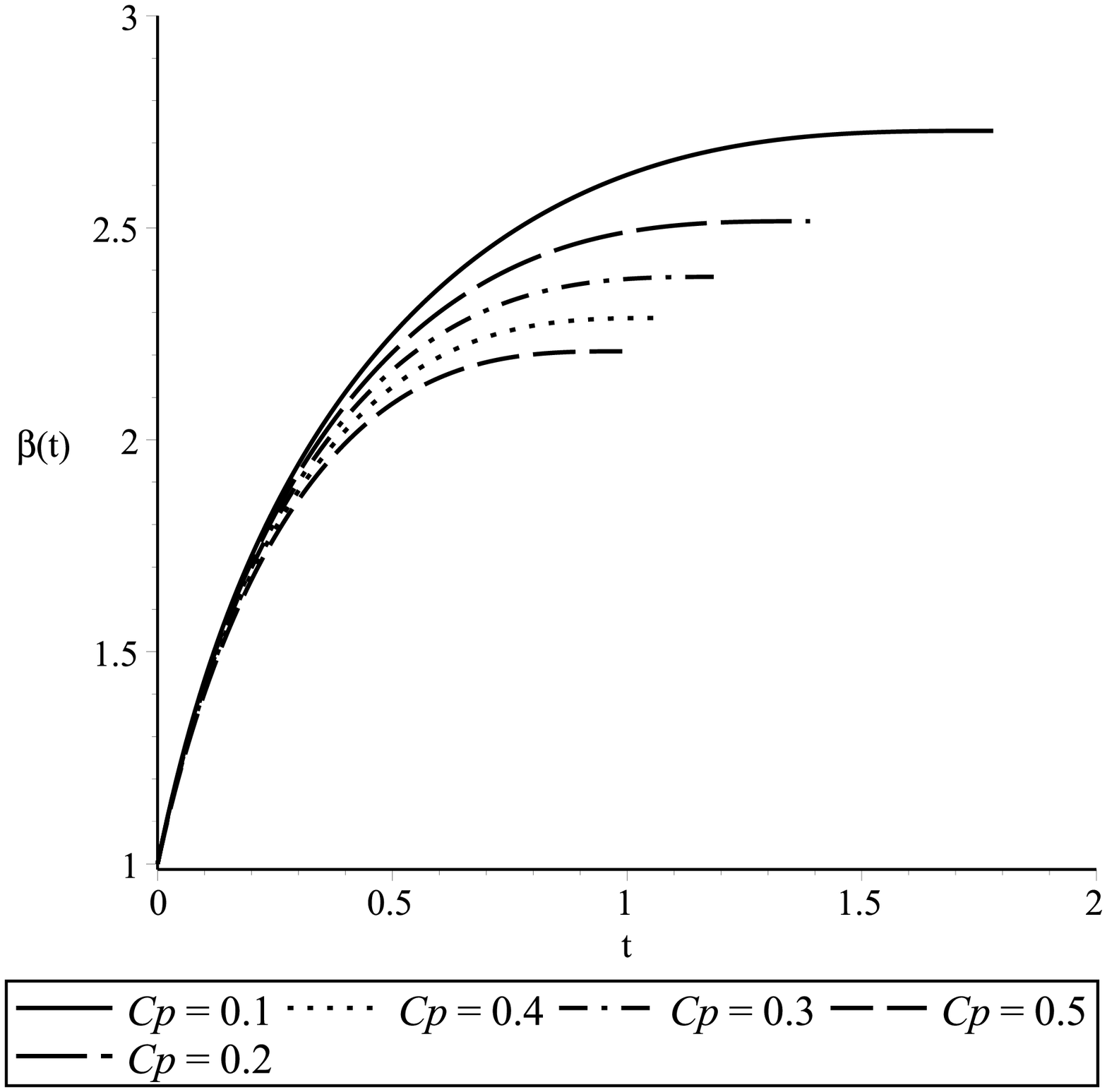}
\end{center}
\caption{$\beta(t)$ as a function of $t$ for different values of $C_p$ with $w = -2$, $C_r = C_d = 1$}
\label{figure_betaCp}
\end{figure}

Consider, now, how the phantom fluid modifies, quantitatively, the dynamical evolution of $a(t)$ and $\beta(t)$. In order to do that, we vary the phantom fluid energy density $C_p$. After computing the solution to the system Eqs. (\ref{10}), (\ref{14}), for many different values of 
$C_p$ with fixed values of $w$, $C_r$ and $C_d$, we find that: the scale factor $a(t)$ expands more rapidly for greater values of $C_p$ and the anisotropy parameter $\beta(t)$ tends, asymptotically, to smaller constant values when one increases the value of $C_p$. We present, respectively, in Figures \ref{figure_aCp} and \ref{figure_betaCp} examples of those behaviors. In Table \ref{T4}, one can see the values of $t_s$ and $a(t)$, $\beta(t)$, $\dot{a}(t)$, $\dot{\beta}(t)$ at $t = t_s$, for the values of $C_p$ shown in Figures \ref{figure_aCp} and \ref{figure_betaCp}. From that Table, it is clear that $\beta(t)$ tends, asymptotically, to a constant value, because its time derivative tends, asymptotically, to zero. Now, we may compare the differences between $C_p$ and the other two energy densities $C_r$ and $C_d$.
For models where $a(t)$ expands more rapidly, for increasing values of $C_p$, $C_r$ or $C_d$, $a(t)$ expands much more rapidly when one increases $C_p$ rather than $C_r$ or $C_d$. We give an example of that behavior in Table \ref{T41}. There, we compare the values of $t_s$ for four models. In the first one, which we call $M1$, we vary $C_r$, leaving the other quantities fixed: $C_d = 100$, $C_p = 1$. In the second one, which we call $M2$, we vary $C_d$, leaving the other quantities fixed: $C_r = 100$, $C_p = 1$. Since, we want to consider the models as similar as possible, in order to facilitate the comparison, we are left with two choices for the other two models. In the third model, which we call $M3$, we vary $C_p$, leaving the other quantities fixed: $C_r = 100$, $C_d = 1$. Finally, in the fourth model, which we call $M4$, we vary $C_p$, leaving the other quantities fixed: $C_r = 1$, $C_d = 100$. In all four models, we consider $w=-2$. In order to obtain a better comparison, we had to modify the initial condition $\dot{a}_0$, in all four models, from its usual value Eq. (\ref{15}), to the new value 30. The values of the other initial conditions are as given in Eq. (\ref{15}). Taking in account the results of that comparison, between $C_p$, $C_r$ and $C_d$, we may say that the phantom fluid gives the most important contribution to the expansion of the universe, in those models.

\begin{table}[!htb]
\centering
\caption{\footnotesize Values of $\beta(t)$, $\dot{\beta}(t)$, $a(t)$ and $\dot{a}(t)$ at $t = t_s$, for
$C_r = C_d = 1$, $w = -2$ and different values of $C_p$.}
\label{T4}
{\scriptsize\begin{tabular}{c c c c c c}
\hline
$C_p$ &  $t_{s}$ & $\beta(t_{s})$ & $\dot{\beta}(t_{s}) \times 10^{-14}$ & $a(t_{s})$ & $\dot{a}(t_{s}) \times 10^{11}$ \\
\hline

 0.1 & 1.7869482 & 2.72867613812881 & $3.17591472402185$ & 416852.662835199 & $405.189398243791$ \\

 0.2 & 1.3932501 & 2.51549234776996 & $26.5910018013945$ & 80871.9031148746 & $9.00668140248792$ \\

 0.3 & 1.2019288 & 2.38427551539225 & $3.40270451509271$ & 281888.414707609 & $242.137832091572$ \\

 0.4 & 1.0813888 & 2.28712172779703 & $14.3430643077579$ & 103970.495744935 & $22.5458059191255$ \\

 0.5 & 0.99585439 & 2.20862349609312 & $3.82346026017136$ & 236223.668097204 & $192.322965295604$ \\

\hline
\end{tabular}}
\end{table}

\begin{table}[!htb]
\centering
\caption{\footnotesize Comparison between $C_r$, $C_d$ and $C_p$ for models where $a(t)$ expands more rapidly for increasing values of $C_r$, 
$C_d$ or $C_p$.}
\label{T41}
{\scriptsize\begin{tabular}{c c c c c}
\hline
$C_r(M1)/C_d(M2)/C_p(M3, M4)$ & $t_s(M1)$ & $t_s(M2)$ & $t_s(M3)$ & $t_s(M4)$\\
\hline
			
100  &  0.28282753 & 0.28282753 & 0.057496990 & 0.057448376\\
			
200  &  0.28223448 & 0.28043677 & 0.044671526 & 0.044646024\\
			
300  &  0.28165043 & 0.27815985 & 0.038468839 & 0.038451755\\

400  &  0.28107516 & 0.27598714 & 0.034570618 & 0.034557949\\

500  &  0.28050850 & 0.27391022 & 0.031808746 & 0.031798812\\
			
\hline
\end{tabular}}
\end{table}

\subsection{Varying $a_0$}

\begin{figure}
\begin{center}
\includegraphics[height=7.5cm,width=9.5cm]{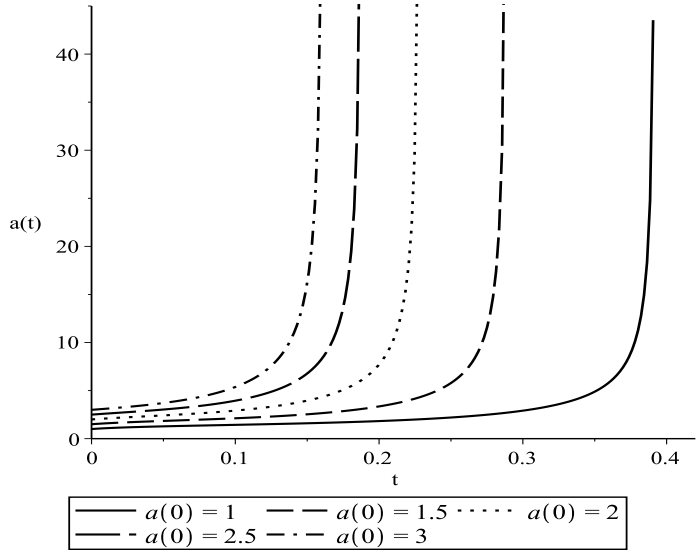}
\end{center}
\caption{$a(t)$ as a function of $t$ for different values of $a_0$ with $w = -2$, $C_r = C_d = C_p = 1$}
\label{figure_aa0}
\end{figure}

\begin{figure}
\begin{center}
\includegraphics[height=7.5cm,width=9.5cm]{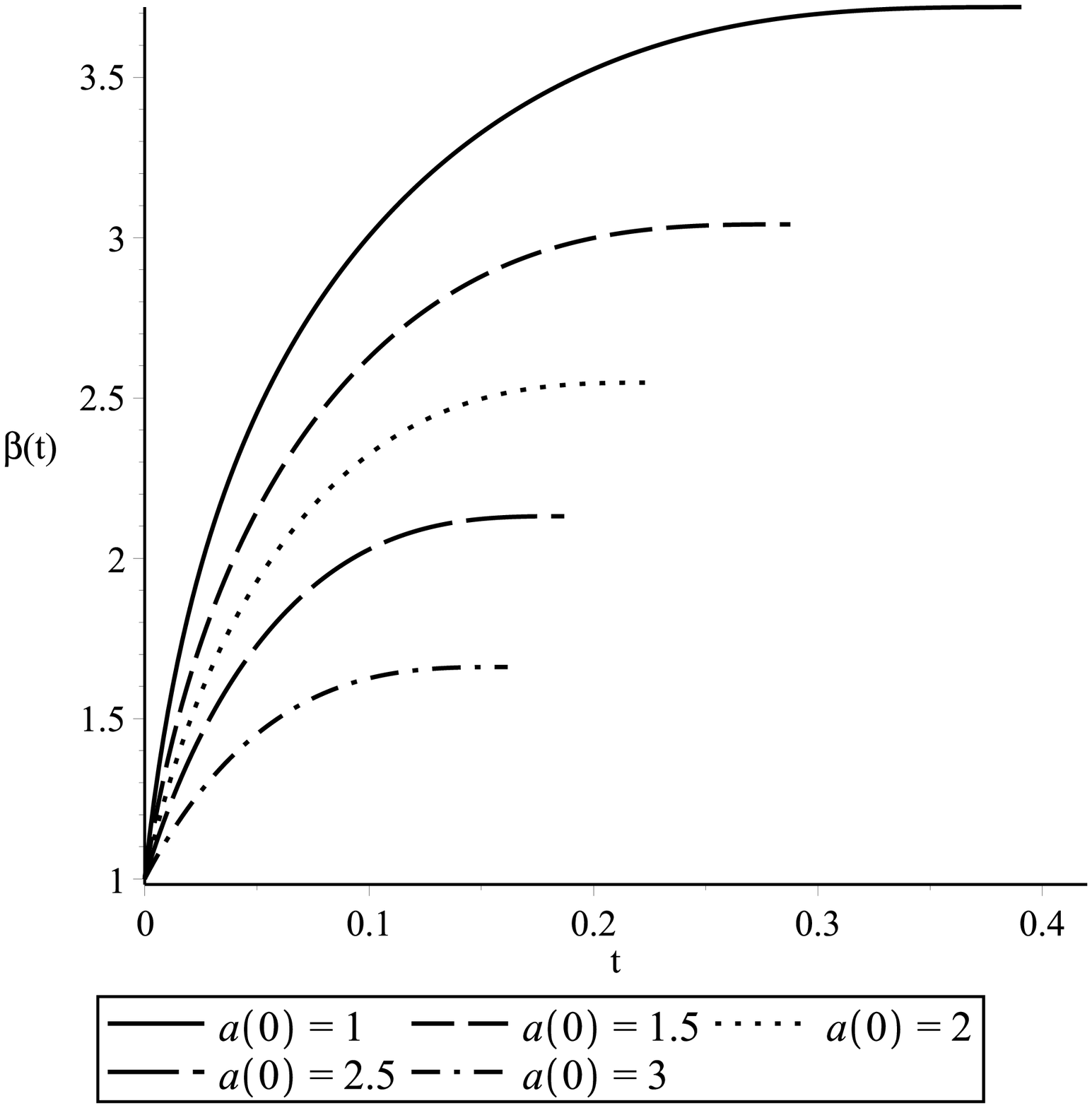}
\end{center}
\caption{$\beta(t)$ as a function of $t$ for different values of $a_0$ with $w = -2$, $C_r = C_d = C_p = 1$}
\label{figure_betaa0}
\end{figure}

We start studying how the initial conditions modify, quantitatively, the dynamical evolution of $a(t)$ and $\beta(t)$. We start considering the scale factor initial condition $a_0$. After computing the solution to the system Eqs. (\ref{10}), (\ref{14}), for many different values of 
$a_0$ with fixed values of $w$, $C_r$, $C_d$ and $C_p$, we find that: the scale factor $a(t)$ expands more rapidly for greater values of $a_0$ and the anisotropy parameter $\beta(t)$ tends, asymptotically, to smaller constant values when one increases the value of $a_0$. We present, respectively, in Figures \ref{figure_aa0} and \ref{figure_betaa0} examples of these behaviors. In Table \ref{T5}, one can see the values of $t_s$ and $a(t)$, $\beta(t)$, $\dot{a}(t)$, $\dot{\beta}(t)$ at $t = t_s$, for the values of $a_0$ shown in Figures \ref{figure_aa0} and \ref{figure_betaa0}. From that Table, it is clear that $\beta(t)$ tends, asymptotically, to a constant value, because its time derivative tends, asymptotically, to zero. In order to obtain a better example of the present case, we had to modify the initial condition 
$\dot{a}_0$, from its usual value Eq. (\ref{15}), to the new value 11. The values of the other initial conditions are as given in Eq. (\ref{15}).

\begin{table}[!htb]
\centering
\caption{\footnotesize Values of $\beta(t)$, $\dot{\beta}(t)$, $a(t)$ and $\dot{a}(t)$ at $t = t_s$, for $C_r = C_d = C_p = 1$, $w = -2$ and different values of $a_0$.}
\label{T5}
{\scriptsize\begin{tabular}{c c c c c c}
\hline
$a_0$ &  $t_{s}$ & $\beta(t_{s})$ & $\dot{\beta}(t_{s}) \times 10^{-15}$ & $a(t_{s})$ & $\dot{a}(t_{s}) \times 10^{13}$ \\
\hline

 1.0 & 0.39218666 & 3.71911243894487 & $41.1325082986648$ & 147023.853697808 & $1.21256917353076$ \\

 1.5 & 0.28881554 & 3.04144094545045 & $5.43630247148688$ & 593717.667479622 & $33.5436037875048$ \\

 2.0 & 0.22852645 & 2.54783078445521 & $8.34979507088098$ & 504963.110391576 & $19.7795253185517$ \\

 2.5 & 0.18850147 & 2.13104958098695 & $30.3700304317329$ & 275475.887662819 & $3.91763753194575$ \\

 3.0 & 0.16170791 & 1.66066972121717 & $53.9125523434728$ & 227090.776172289 & $2.14903323073259$ \\

\hline
\end{tabular}}
\end{table}

\subsection{Varying $\beta_0$}

\begin{figure}
\begin{center}
\includegraphics[height=7.5cm,width=9.5cm]{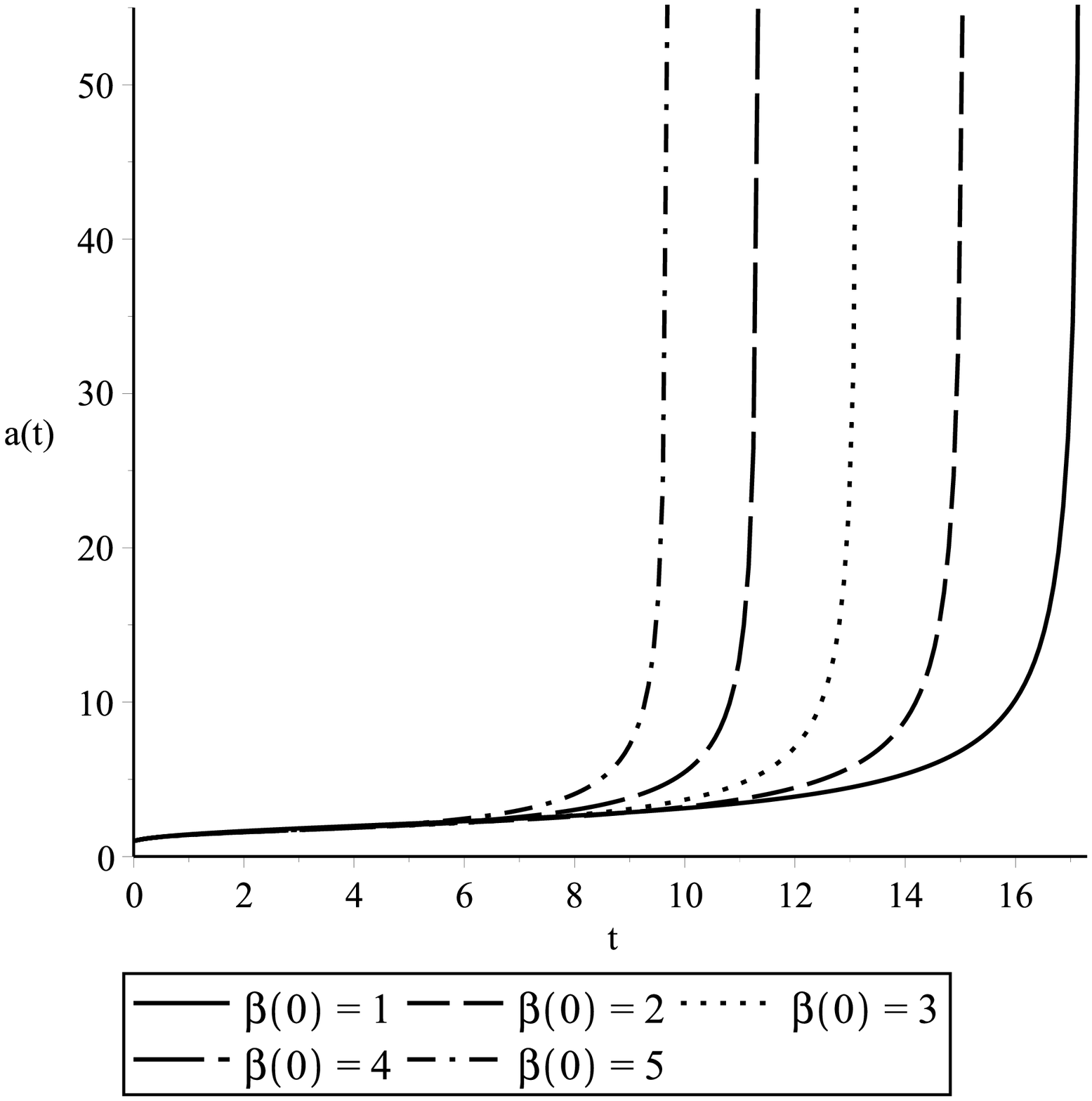}
\end{center}
\caption{$a(t)$ as a function of $t$ for different values of $\beta_0$ with $w = -2$, $C_r = C_d = 1$, $C_p = 0.0001$}
\label{figure_abeta0}
\end{figure}

\begin{figure}
\begin{center}
\includegraphics[height=7.5cm,width=9.5cm]{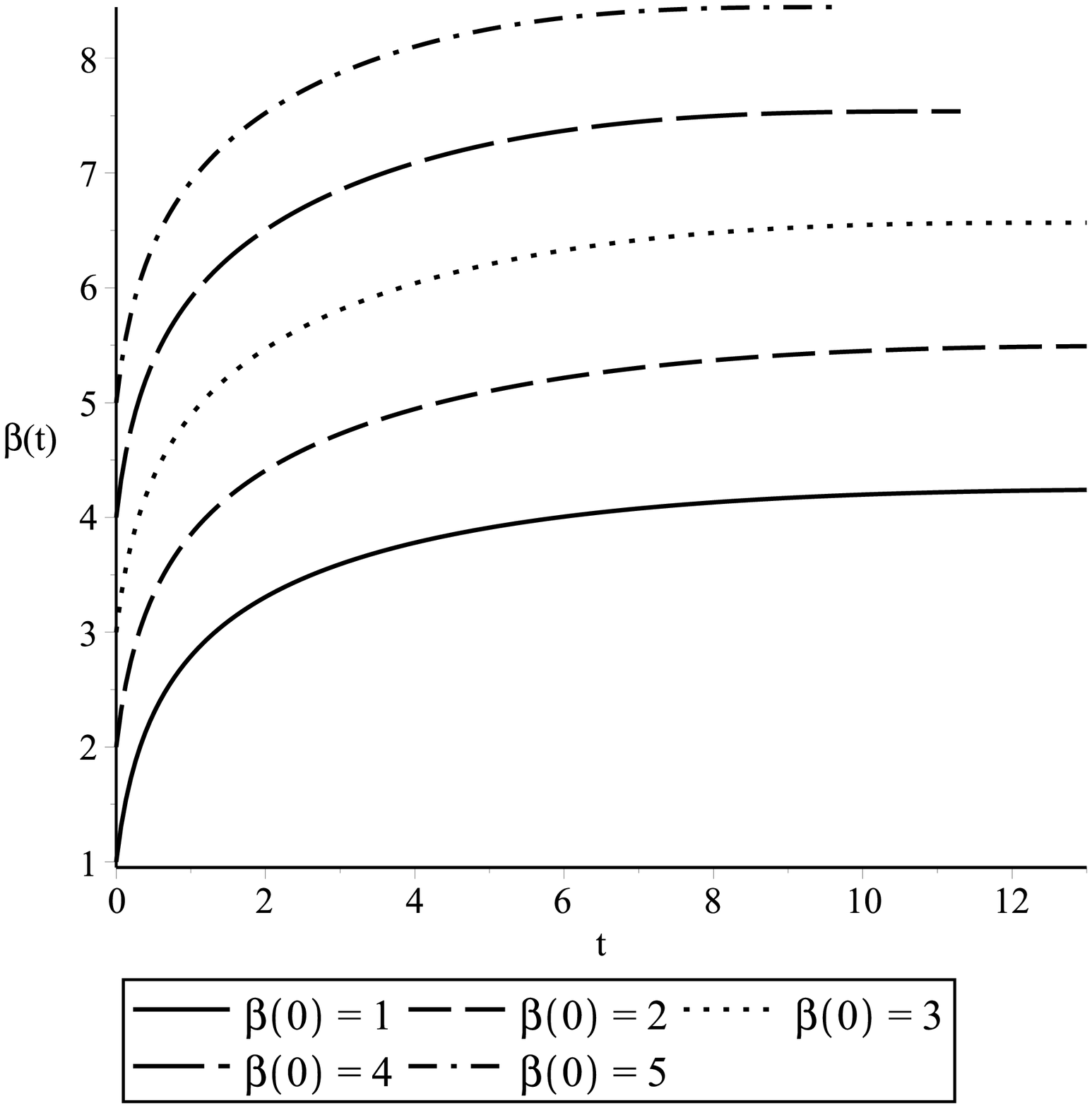}
\end{center}
\caption{$\beta(t)$ as a function of $t$ for different values of $\beta_0$ with $w = -2$, $C_r = C_d = 1$, $C_p = 0.0001$}
\label{figure_betabeta0}
\end{figure}

Let us consider, how $\beta_0$ modifies, quantitatively, the dynamical evolution of $a(t)$ and $\beta(t)$. After computing the solution to the system Eqs. (\ref{10}), (\ref{14}), for many different values of $\beta_0$ with fixed values of $w$, $C_r$, $C_d$ and $C_p$, we find that: the scale factor $a(t)$ expands more rapidly for greater values of $\beta_0$ and the anisotropy parameter $\beta(t)$ tends, asymptotically, to greater constant values, when one increases the value of $\beta_0$. We present, respectively, in Figures \ref{figure_abeta0} and \ref{figure_betabeta0} examples of these behaviors. In Table \ref{T6}, one can see the values of $t_s$ and $a(t)$, $\beta(t)$, $\dot{a}(t)$, $\dot{\beta}(t)$ at $t = t_s$, for the values of $\beta_0$ shown in Figures \ref{figure_abeta0} and \ref{figure_betabeta0}. From that Table, it is clear that $\beta(t)$ tends, asymptotically, to a constant value, because its time derivative tends, asymptotically, to zero. 

\begin{table}[!htb]
\centering
\caption{\footnotesize Values of $\beta(t)$, $\dot{\beta}(t)$, $a(t)$ and $\dot{a}(t)$ at $t = t_s$, for $C_r = C_d = 1$, $C_p = 0.0001$, $w = -2$ and different values of $\beta_0$.}
\label{T6}
{\scriptsize\begin{tabular}{c c c c c c}
\hline
$\beta_0$ & $t_{s}$ & $\beta(t_{s})$ & $\dot{\beta}(t_{s}) \times 10^{-13}$ & $a(t_{s})$ & $\dot{a}(t_{s}) \times 10^{10}$ \\
\hline

 1.0 & 17.236542 & 4.24714293260463 & $2.55148791589835$ & 793419.119050158 & $936.103343315701$ \\

 2.0 & 15.117073 & 5.49468110851415 & $40.7394949220433$ & 99472.0360334776 & $7.11654930852383$ \\

 3.0 & 13.186692 & 6.56749755714504 & $35.3402974720603$ & 91480.9013729014 & $7.54783863833558$ \\

 4.0 & 11.374163 & 7.53742056652959 & $35.7124854815463$ & 80145.6074495621 & $6.91040870203405$ \\

 5.0 & 9.7209844 & 8.44547655097989 & $1.30522919915676$ & 627531.502714957 & $1487.59315119145$ \\

\hline
\end{tabular}}
\end{table}

\subsection{Varying $\dot{a}_0$}

\begin{figure}
\begin{center}
\includegraphics[height=7.5cm,width=9.5cm]{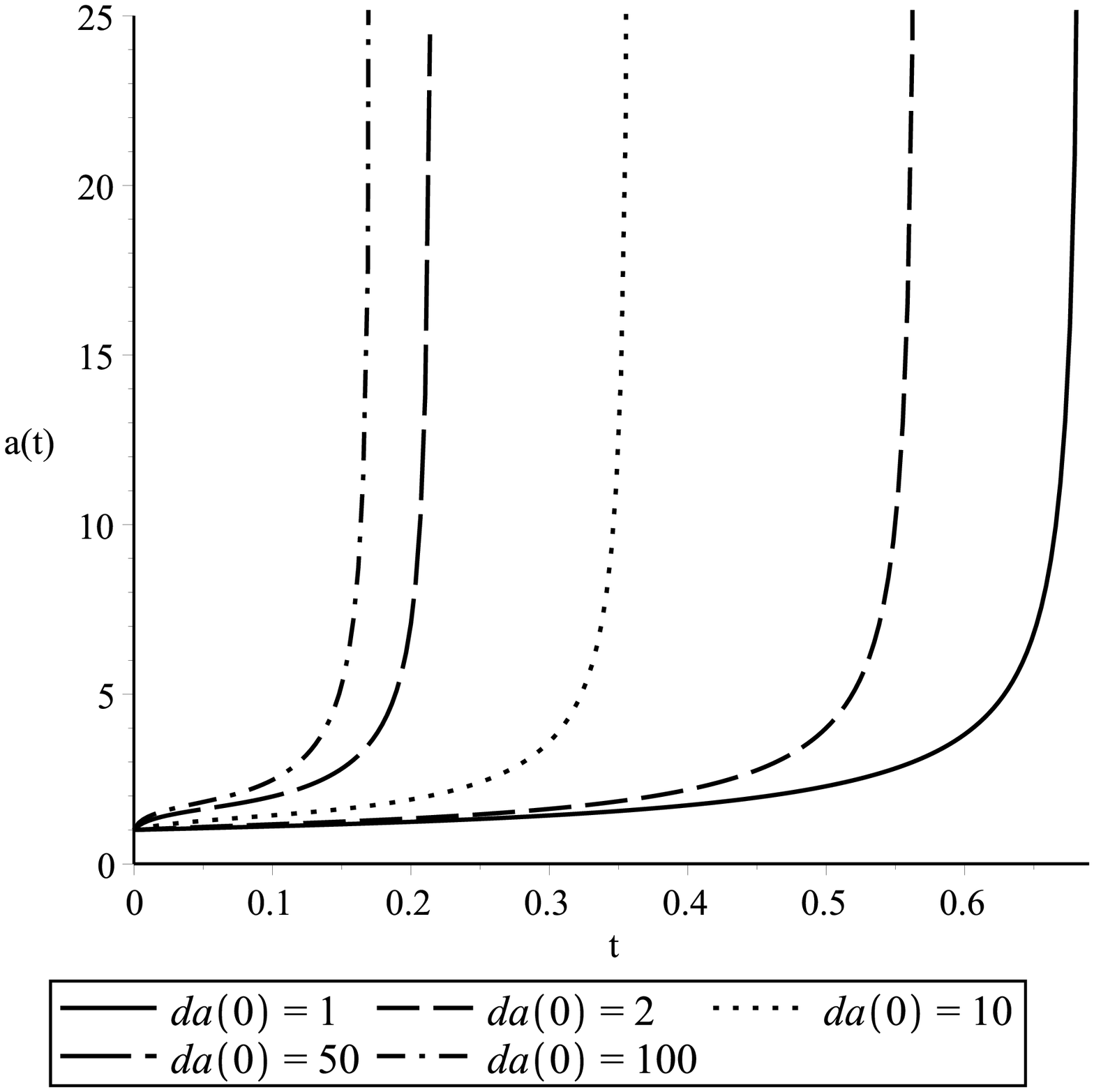}
\end{center}
\caption{$a(t)$ as a function of $t$ for different values of $\dot{a}_0$ with $w = -2$, $C_r = C_d = C_p = 1.4$}
\label{figure_ada0}
\end{figure}

\begin{figure}
\begin{center}
\includegraphics[height=7.5cm,width=9.5cm]{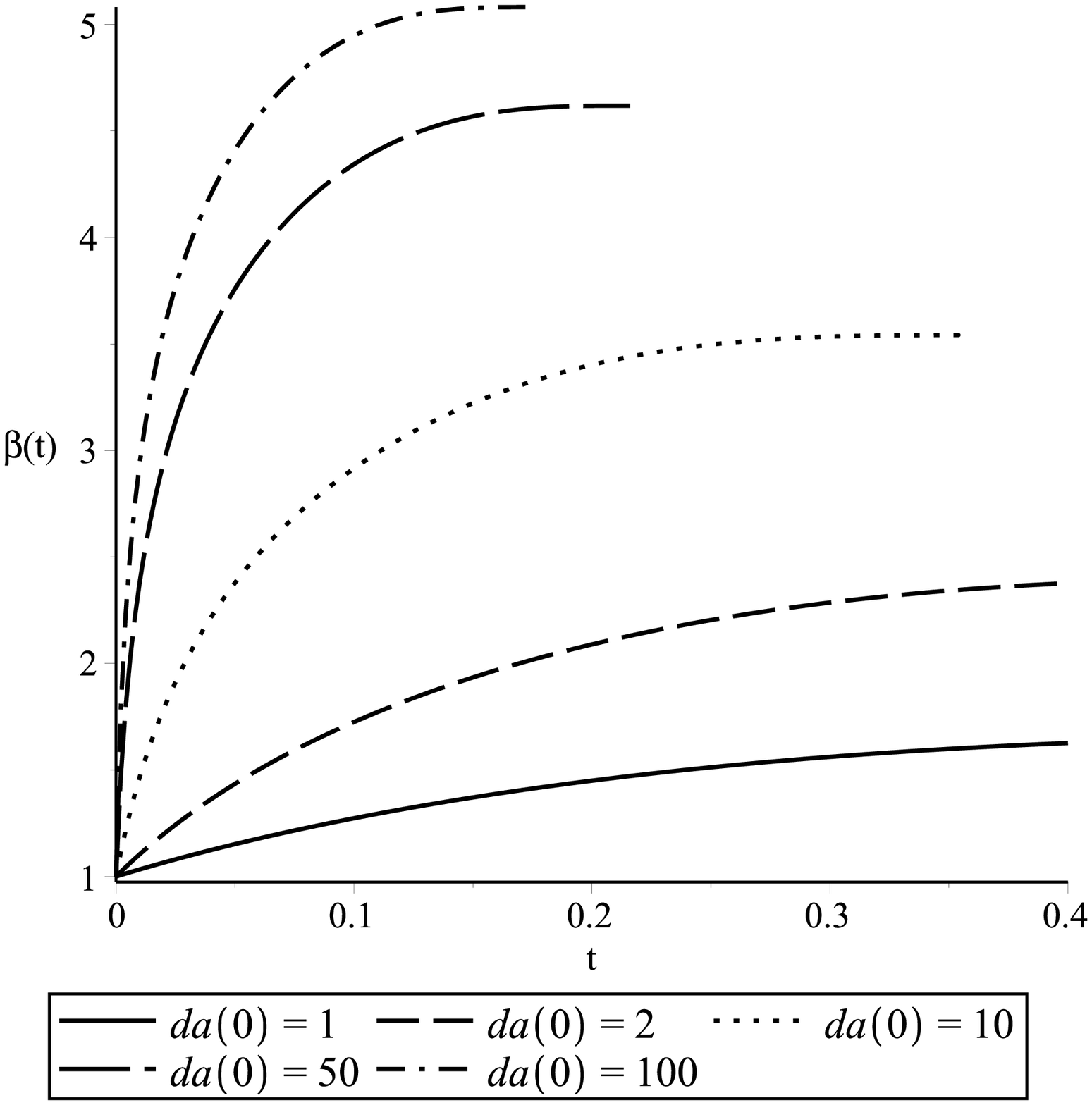}
\end{center}
\caption{$\beta(t)$ as a function of $t$ for different values of $\dot{a}_0$ with $w = -2$, $C_r = C_d = C_p = 1.4$}
\label{figure_betada0}
\end{figure}

The next initial condition we consider is $\dot{a}_0$. Let us study how it modifies, quantitatively, the dynamical evolution of $a(t)$ and $\beta(t)$. After computing the solution to the system Eqs. (\ref{10}), (\ref{14}), for many different values of $\dot{a}_0$ with fixed values of $w$, $C_r$, $C_d$ and $C_p$, we find that: the scale factor $a(t)$ expands more rapidly for greater values of $\dot{a}_0$ and the anisotropy parameter $\beta(t)$ tends, asymptotically, to greater constant values, when one increases the value of $\dot{a}_0$. We present, respectively, in Figures \ref{figure_ada0} and \ref{figure_betada0} examples of these behaviors. In Table \ref{T7}, one can see the values of $t_s$ and $a(t)$, $\beta(t)$, $\dot{a}(t)$, $\dot{\beta}(t)$ at $t = t_s$, for the values of $\beta_0$ shown in Figures \ref{figure_ada0} and \ref{figure_betada0}. From that Table, it is clear that $\beta(t)$ tends, asymptotically, to a constant value, because its time derivative tends, asymptotically, to zero.

\begin{table}[!htb]
\centering
\caption{\footnotesize Values of $\beta(t)$, $\dot{\beta}(t)$, $a(t)$ and $\dot{a}(t)$ at $t = t_s$, for $C_r = C_d = C_p = 1.4$, $w = -2$ and different values of $\dot{a}_0$.}
\label{T7}
{\scriptsize\begin{tabular}{c c c c c c}
\hline
$\dot{a}_0$ & $t_{s}$ & $\beta(t_{s})$ & $\dot{\beta}(t_{s}) \times 10^{-14}$ & $a(t_{s})$ & $\dot{a}(t_{s}) \times 10^{12}$ \\
\hline

 1.0 & 0.68637653 & 1.66977494354006 & $1.59090705364705$ & 320185.469370538 & $60.1590639209326$ \\

 2.0 & 0.56685504 & 2.40606227176932 & $4.73076060356115$ & 142553.373933854 & $9.56493558847178$ \\

 10.0 & 0.35925246 & 3.54167195387590 & $1.32259169459675$ & 278362.335418070 & $67.6958086970405$ \\

 50.0 & 0.21644327 & 4.61907753488693 & $8.68319955623955$ & 102943.433066229 & $7.37072199231872$ \\

 100.0 & 0.17277255 & 5.08146132578450 & $14.6361427772638$ & 91174.2212218933 & $6.10796228056579$ \\

\hline
\end{tabular}}
\end{table}

\subsection{Varying $\dot{\beta}_0$}

\begin{figure}
\begin{center}
\includegraphics[height=7.5cm,width=9.5cm]{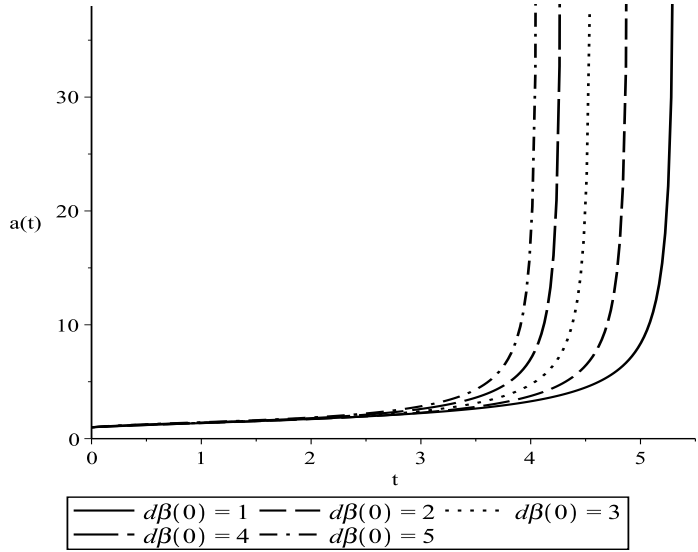}
\end{center}
\caption{$a(t)$ as a function of $t$ for different values of $\dot{\beta}_0$ with $w = -2$, $C_r = C_d = 1$ and $C_p = 0.01$}
\label{figure_adbeta0}
\end{figure}

\begin{figure}
\begin{center}
\includegraphics[height=7.5cm,width=9.5cm]{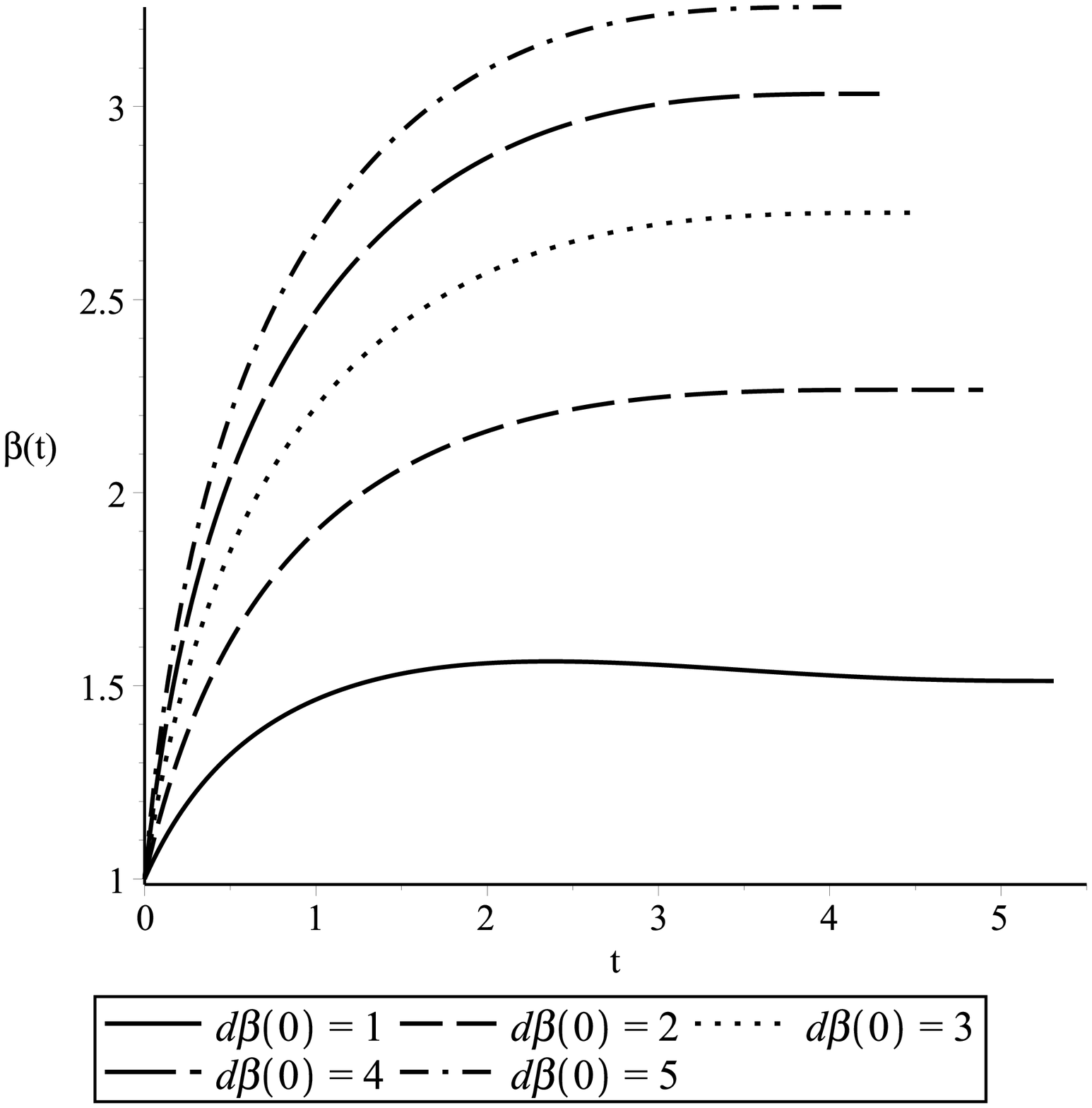}
\end{center}
\caption{$\beta(t)$ as a function of $t$ for different values of $\dot{\beta}_0$ with $w = -2$, $C_r = C_d = 1$ and $C_p = 0.01$}
\label{figure_betadbeta0}
\end{figure}

Finally, we consider how $\dot{\beta}_0$ modifies, quantitatively, the dynamical evolution of $a(t)$ and $\beta(t)$. After computing the solution to the system Eqs. (\ref{10}), (\ref{14}), for many different values of $\dot{\beta}_0$ with fixed values of $w$, $C_r$, $C_d$ and $C_p$, we find that: the scale factor $a(t)$ expands more rapidly for greater values of $\dot{\beta}_0$ and the anisotropy parameter $\beta(t)$ tends, asymptotically, to greater constant values, when one increases the value of $\dot{\beta}_0$. We present, respectively, in Figures \ref{figure_adbeta0} and \ref{figure_betadbeta0} examples of these behaviors. In Table \ref{T8}, one can see the values of $t_s$ and $a(t)$, $\beta(t)$, $\dot{a}(t)$, $\dot{\beta}(t)$ at $t = t_s$, just before reaching the singularity, for the values of $\dot{\beta}_0$ shown in Figures \ref{figure_adbeta0} and \ref{figure_betadbeta0}. From that Table, it is clear that $\beta(t)$ tends, asymptotically, to a constant value, because its time derivative tends, asymptotically, to zero. In this case, we let $\dot{a}_0$ vary freely and determined its values using the first order, ordinary differential equation (\ref{10}), for the initial conditions. The values of the other initial conditions are as given in Eq. (\ref{15}).

\begin{table}[!htb]
\centering
\caption{\footnotesize Values of $\beta(t)$, $\dot{\beta}(t)$, $a(t)$ and $\dot{a}(t)$ at $t = t_s$, for $C_r = C_d = 1$, $C_p = 0.01$, $w = -2$ and different values of $\dot{\beta}_0$.}
\label{T8}
{\scriptsize\begin{tabular}{c c c c c c}
\hline
$\dot{\beta}_0$ & $t_{s}$ & $\beta(t_{s})$ & $\dot{\beta}(t_{s}) \times 10^{-14}$ & $a(t_{s})$ & $\dot{a}(t_{s}) \times 10^{12}$ \\
\hline

 1.0 & 5.3281792 & 1.51238420873302 & $9.53849637867111$ & 515855.950529562 & $16.1051818520860$ \\

 2.0 & 4.9028128 & 2.26631291462971 & $5.40273034724179$ & 667353.915342209 & $37.0161974672880$ \\

 3.0 & 4.5628473 & 2.72500139914146 & $2.04926752114798$ & 1189030.53897112 & $175.907851098471$ \\

 4.0 & 4.2914049 & 3.03296489052531 & $20.9437719271073$ & 236098.831823163 & $3.33790542900046$ \\

 5.0 & 4.0705723 & 3.25739822988567 & $8.14521315626682$ & 422511.134085790 & $15.1252454938635$ \\

\hline
\end{tabular}}
\end{table}

\section{Conclusions}
\label{conclusions}

In the present work, we studied the dynamical evolution of an homogeneous and anisotropic KS cosmological model, considering general relativity as the gravitational theory, such that there are three different perfect fluids in the matter sector. They are radiation, dust and phantom fluid. Our main motivation was verifying if the present model tends to an homogeneous and isotropic FRW model, during its evolution. Also, we wanted to establish how the parameters and initial conditions of the model, quantitatively, influence the isotropization of the present model. In order to simplify our task, we used the Misner parametrization of the KS metric. In terms of that parametrization the KS metric has two metric functions: the scale factor $a(t)$ and $\beta(t)$, which measures the spatial anisotropy of the model. In terms of
those metric functions, the isotropization takes place if $\beta(t)$ tends to zero or a constant value, for a given, finite or infinity, value of time.
Initially, we obtained the Hamiltonian of the model. Then, from that Hamiltonian, we computed the three, independent, Einstein's equations. Before we solved them, we identified the two different types of solutions, drawing some phase portraits of the model. The first type of solution is expansive and after some time the isotropization, of the model, takes place. The second type of solution is not expansive. There, after an initial period of expansion $a(t)$ and $\beta(t)$ reach their maximum values and, finally, they contract to their initial values. In the second type of solution, we do not have isotropization of the model. 
After that qualitative study of the solutions, we started solving, numerically, a system formed by two second order, ordinary, differential equations. They were obtained from the Einstein's equations. We restricted our attention to the study of the expansive solutions. We investigated, in details, how the phantom fluid parameter $w$, the energy density parameters of the fluids $C_r$, $C_d$, $C_p$ and the initial conditions $a_0$, $\beta_0$, $\dot{a}_0$, $\dot{\beta}_0$, modify the evolution of $a(t)$ and $\beta(t)$. 

As the initial result of our investigations, we determined that, for all solutions of the expansive type, $a(t)$ starts to expand from a small finite value and after a finite time interval it reaches an infinite value giving rise to a \textit{Big Rip} singularity. If we want that the present model describes our Universe, that period of rapid accelerated expansion, which, we believe, took place when the Universe was very young, must had finished before the development of the \textit{Big Rip} singularity. It means that, after the Universe had expanded, in an accelerated rate, for some time and had developed an high level of isotropy, it started to expand in a slower rate and did not form the  \textit{Big Rip} singularity. Then, after another period of time, it started to expand in the present way. We, also, determined that $\beta(t)$ starts to expand from a small finite value and then it goes, asymptotically, to a constant, when the universe goes to the \textit{Big Rip} singularity. The fact that, $\beta(t)$ goes, asymptotically, to a constant is very important, because it guarantees that the solution is asymptotically isotropic. More precisely, the behavior of $\beta(t)$ means that the KS metric (\ref{4}), goes, asymptotically, to a FRW metric, where $a(t)$ plays the role of a scale factor. We determined that $\beta(t)$ goes to a constant by computing the value of its time derivative ($\dot{\beta}(t)$), for different moments during its evolution. We observed that $\dot{\beta}(t)$ decreases for increasing values of the time coordinate $t$, such that in the last moment before the universe reaches the \textit{Big Rip} singularity ($t_s$), that quantity is very small.
Finally, from our investigations of the expansive solutions, we concluded that: (i) $a(t)$ expands more rapidly for greater values of $C_p$ and all the initial conditions; (ii) $a(t)$ expands more rapidly for smaller values of $w$; (iii) $a(t)$ expands more rapidly or slowly, depending upon certain conditions, for greater values of $C_r$ or $C_d$; (iv) $\beta(t)$ tends, asymptotically, to greater constant values, when one increases the values of $w$, $\beta_0$, $\dot{a}_0$, $\dot{\beta}_0$; (v) $\beta(t)$ tends, asymptotically, to smaller constant values, when one increases the values of $C_r$, $C_d$, $C_p$, $a_0$. Based on our results, we can speculate that the expansive solution may represent an initial, anisotropic, stage of our Universe, that later, due to the expansion, became isotropic.

{\bf Acknowledgments}. D. L. Canedo thanks Coordena\c{c}\~{a}o de Aperfei\c{c}oamento de Pessoal de N\'{i}vel Superior (CAPES) and Universidade Federal de Juiz de Fora (UFJF) for his scholarships. G. A. Monerat thanks FAPERJ for financial support (Proc. E-26/010.101230/2018).


\begin{thebibliography}{00}

\bibitem{wheeler} C. W. Misner, K. S. Thorne and J. A. Wheeler, {\it Gravitation}, (W. H. Freeman and Company, New York, 1973).

\bibitem{dinverno} R. D'Inverno, {\it Introducing Einstein's Relativity}, (Oxford University Press, Oxford, 1998).

\bibitem{liddle} A. R. Liddle, {\it An Introduction to Modern Cosmology}, (Wiley \& Sons, Chichester, 2015).

\bibitem{liddle1} A. R. Liddle and D. H. Lyth, {\it Cosmological Inflation and Large-Scale Structure}, (Cambridge University Press, Cambridge, 2000).

\bibitem{misner0} See for instance: C. W. Misner, Astrophys. J. {\bf 151}, p. 431 (1968).

\bibitem{KS} R. Kantowski and R. K. Sachs, J. Math. Phys. {\bf 7}, p. 443 (1966).

\bibitem{lorenz} D. Lorenz, J. Phys. A: Math. Gen. {\bf 16}, pp. 575-584 (1983).

\bibitem{weber} E. Weber, J. Math. Phys. {\bf 25}, p. 3279 (1984).

\bibitem{weber1} E. Weber, J. Math. Phys. {\bf 26}, p. 1308 (1985).

\bibitem{gron} \O . Gr\o n, J. Math. Phys. {\bf 27}, p. 1490 (1986).

\bibitem{gron1} \O . Gr\o n and E. Eriksen, Phys. Lett. A {\bf 121}, p. 217 (1987).

\bibitem{barrow} A. B. Burd and J. D. Barrow, Nucl. Phys. B. {\bf 308}, pp. 929-945 (1988).

\bibitem{jensen} L. G. Jensen and P. J. Ruback, Nucl. Phys. B. {\bf 325}, pp. 660-686 (1989).

\bibitem{krori} K. D. Krori, J. Math. Phys. {\bf 36}, p. 1347 (1995).

\bibitem{byland} S. Byland and D. Scialom, Phys. Rev. D {\bf 57}, p. 6065 (1998).

\bibitem{li} X. Z. Li and J. G. Hao, Phys. Rev. D {\bf 68}, 083512 (2003).

\bibitem{tiwari} R. K. Tiwari and U. K. Dwivedi, Astrophys. Space Sci. {\bf 318}, pp. 249-253 (2008).

\bibitem{adhav} K. S. Adhav, A. S. Bansod, R. P. Wankhade and H. G. Ajmire, Cent. Eur. J. Phys. {\bf 9}, pp. 919-925 (2011).

\bibitem{chaubey} R. Chaubey, Int. J. Astron. and Astrophys. {\bf 1}, pp. 25-38 (2011).

\bibitem{adhav1} K. S. Adhav, Eur. Phys. J. Plus {\bf 126}, 103 (2011).

\bibitem{parisi} L. Parisi, N. Radicella and G. Vilasi, Phys. Rev. D {\bf 91}, 063533 (2015).

\bibitem{keresztes} Z. Keresztes, M. Forsberg, M. Bradley, P. K. S. Dunsby and L. \'A. Gergely, JCAP {\bf 11}, 042 (2015).

\bibitem{singh} T. Singh and A. K. Agrawal, Astrophys. Space Sci. {\bf 182}, pp. 289-312 (1991).

\bibitem{barrow1} J. D. Barrow and M. P. Dabrowski, Phys. Rev. D {\bf 55}, 630 (1997).

\bibitem{samanta} G. C. Samanta, Int. J. Theor. Phys. {\bf 52}, pp. 2647-2656 (2013).

\bibitem{latta} J. Latta, G. Leon and A. Paliathanasis, JCAP {\bf 11}, 051 (2016).

\bibitem{dutta} S. Dutta, M. Lakshmanan, S. Chakraborty, Annals Phys. {\bf 393}, pp. 254-263 (2018).

\bibitem{vinutha} T. Vinutha, V. U. M. Rao, B. Getaneh and M. Mengesha, Astrophys. Space Sci. {\bf 363}, 188 (2018).

\bibitem{hoogen} R. J. van den Hoogen {\it et al}, JCAP {\bf 11}, 017 (2018).

\bibitem{mishra} S. Mishra and S. Chakraborty, Annals Phys. {\bf 406}, pp. 207-219 (2019).

\bibitem{cesare} M. Cesare, S. S. Seahra and E. Wilson-Ewing, JCAP {\bf 07}, 018 (2020).

\bibitem{mohandas} S. Mohandas, R. J. van den Hoogen, D. Winters and M. Dala, JCAP {\bf 08}, 021 (2020).

\bibitem{leon} G. Leon, A. Paliathanasis and N. Dimakis, Eur. Phys. J. C {\bf 80}, 1149 (2020).

\bibitem{laflamme} R. Laflamme and E. P. S. Shellard, Phys. Rev. D {\bf 35}, 2315 (1987).

\bibitem{halliwell} J. J. Halliwell and J. Louko, Phys. Rev. D {\bf 42}, 3997 (1990).

\bibitem{compean} H. G. Compe\'{a}n, O. Obreg\'{o}n and C. Ram\'{\i}rez, Phys. Rev. Lett. {\bf 88}, 161301 (2002).

\bibitem{nelson} G. D. Barbosa and N. Pinto-Neto, Phys. Rev. D {\bf 70}, 103512 (2004).

\bibitem{obregon} O. Obreg\'{o}n and J. A. Preciado, Phys. Rev. D {\bf 86}, 063502 (2012).

\bibitem{riess0} A. G. Riess et al., Astron. J. {\bf 116}, 1009 (1998). 

\bibitem{perlmutter} S. Perlmutter et al., Astrophys. J. {\bf 517}, 565 (1999).

\bibitem{misner} C. Misner, {\it Minisuperspace}, in Magic without magic: John
Archibald Wheeler, Ed. J. R. Klauder, (W. H. Freeman and Company, San Francisco, 1972), Eq. (23) p. 449.

\bibitem{caldwell} R. R. Caldwell, Phys. Lett. B {\bf 545}, pp. 23–29 (2002).



\end{thebibliography}
\end{document}